\documentclass[10pt,emptycopyrightspace]{ewsn-proc}

\renewcommand{\paragraph}[1]{\noindent\textbf{#1 }}

\usepackage{balance}
\usepackage{comment}

\usepackage{cite}

\usepackage[cmex10]{amsmath}
\usepackage[tight,footnotesize]{subfigure}

\usepackage{url}

\usepackage{graphicx}
\usepackage{graphics}
\usepackage{amssymb}
\usepackage{amsmath}
\usepackage{color}
\usepackage{courier}

\usepackage{multirow}
\usepackage{footnote}
\usepackage[bottom]{footmisc}

\usepackage{listings}
\definecolor{dkgreen}{rgb}{0,0.6,0}
\definecolor{gray}{rgb}{0.5,0.5,0.5}
\definecolor{mauve}{rgb}{0.58,0,0.82}

\usepackage[linesnumbered, ruled]{algorithm2e}
\SetKwRepeat{Do}{do}{while}%

\usepackage{tikz}

\usepackage[hang,small,bf]{caption}
\usepackage{tabulary}

\newcommand{\nop}[1]{}

\usepackage{lipsum}
\usepackage{xcolor}
\usepackage{soul}
\sethlcolor{black}
\makeatletter
\newif\if@blind
\@blindtrue 

\if@blind \sethlcolor{black}\else
   
\fi

\usepackage{epstopdf}

\title{RADIUS: A System for Detecting Anomalous Link Quality Degradation in Wireless Sensor Networks} 
	
\numberofauthors{6}

\author{
	\alignauthor
	Songwei Fu\\
	\affaddr{Networked Embedded Systems, University of Duisburg-Essen, Germany}\\
	\alignauthor
	Chia-Yen Shih\\
	\affaddr{Networked Embedded Systems, University of Duisburg-Essen, Germany}\\
	\alignauthor
	Yuming Jiang\\
	\affaddr{Department of Telematics, Norwegian University of Science and Technology(NTNU), Norway}\\
	\and  
	\alignauthor
	Matteo Ceriotti\\
	\affaddr{Networked Embedded Systems, University of Duisburg-Essen, Germany}\\
	\alignauthor
	Xintao Huan\\
	\affaddr{Networked Embedded Systems, University of Duisburg-Essen, Germany}\\
	\alignauthor
	Pedro Jos\'{e} Marr\'{o}n\\
	\affaddr{Networked Embedded Systems, University of Duisburg-Essen, Germany}\\
	\and
}

\begin{document}
	
\maketitle

\begin{abstract}


To ensure proper functioning of a Wireless Sensor Network (WSN), it is crucial that the network is able to detect anomalies in communication quality (e.g., RSSI), which may cause  performance degradation, so that the network can react accordingly. In this paper, we introduce RADIUS, a lightweight system for the purpose. The design of RADIUS is aimed at minimizing the detection error (caused by normal randomness of RSSI) in discriminating good links from weak links and at reaching high detection accuracy under diverse link conditions and dynamic environment changes. Central to the design is a threshold-based decision approach that has its foundation on the Bayes decision theory. In RADIUS, various techniques are developed to address challenges inherent in applying this approach. In addition, through extensive experiments, proper configuration of the parameters involved in these techniques is identified for an indoor environment. In a prototype implementation of the RADIUS system deployed in an indoor testbed, the results show that RADIUS is accurate in detecting anomalous link quality degradation for all links across the network, maintaining a stable error rate of 6.13\% on average. 

\end{abstract}

\section{Introduction}\label{sec:intro}

The performance of a Wireless Sensor Network (WSN) often deteriorates after in-situ deployment of the network \cite{1182885, 4408504, 6850017, 6661323}. Link quality degradation, due to, e.g., fading and interference, is one of the major reported causes behind such behavior, which may be significant enough to impact the link's performance, e.g., the packet delivery ratio. Detecting such anomalous degradation in link quality is crucial for an operational WSN to decide possible remedy actions such as tuning stack parameters \cite{Lin:atpc, 7164923}. In such way, the network can continuously maintain its performance and satisfy the user's requirements. 

%


In resource constrained WSNs, detecting anomalous link quality degradation requires {\bf lightweight} solutions with low overheads in using memory, computation and communication resources. Resource-hungry centralized monitoring systems \cite{6661323, 1367278, 1267061} and/or machine learning-based detection techniques \cite{4085803, 4289308, 5356174} are hence hardly applicable to WSNs, due to large communication and/or computation overheads. In addition, a solution should be {\bf accurate} with a low error rate (false positive/negative rate) and be {\bf robust} with consistent performance under diverse link conditions and dynamic environment changes. However, in WSNs, due to the stochastic nature of link quality metrics, e.g., received signal strength indicator (RSSI) \cite{2893729}, it is challenging to distinguish between true link quality degradation and normal randomness. Data smoothing\cite{6199865} may only mitigate the problem. CDF-based \cite{4068315, 6199865} and Chebyshev inequality-based \cite{1689248, 1592596, 1515559} statistical techniques are lightweight and seem to be effective in making the distinguishing. However, our investigation, as to be shown later in this paper, reveals that it is difficult to optimize them to achieve both high detection accuracy and robustness for links which may experience diverse link conditions and dynamic environment changes. 

To meet these requirements, i.e., lightweight, accurate and robust, we have designed a system for detecting anomalous link quality degradation, called RADIUS. In addition to being lightweight, its design has also been aimed at minimizing the detection error (caused by normal randomness of RSSI) in discriminating good links from weak links and at being robust in maintaining the detection performance for different links and under dynamic environment changes. Central to the design is a threshold-based decision approach (for being lightweight) that has its foundation on the Bayes decision theory (for being accurate and robust). 

To the best of our knowledge, no prior work has investigated the applicability of Bayesian thresholding in detecting anomalous link quality degradation in WSNs. A possible reason is perhaps due to the various challenges inherent in applying the approach. To address these challenges, various techniques have been developed to identify the number of RSSI samples needed to achieve a ``good'' approximation of the mean and the standard deviation, to update the mean and standard deviation estimates, and to choose and update a ``proper'' setting for the \textit{a priori} probability, where the mean, the standard deviation and the \textit{a priori} probability are the three fundamental variables used in the Bayes formula.


A prototype of the RADIUS system has been implemented and deployed in an indoor testbed. For proper configuration of the parameters involved in the various techniques in RADIUS, suggestions on their settings are given based on extensive experiments. In addition, we found that high detection accuracy can be achieved by RADIUS under diverse link conditions more robustly as compared to the CDF and Chebyshev thresholding techniques. Moreover, the overhead analysis and the detection results show that RADIUS not only has low overheads in memory, communication and computation, but also is accurate in detecting link quality anomalies for all links across the network, maintaining a stable error rate of 6.13\% on average. These are an indication of RADIUS in fulfilling the requirements.

The rest of the paper is organized as follows. Section \ref{sec:related} discusses the related work. Section \ref{sec:system} presents the system design and motivates the adoption of Bayesian thresholding. Section \ref{sec:approach} introduces the key techniques used in RADIUS. Section \ref{sec:parameterChoice} analyzes the effect of the various involved parameters in these RADIUS techniques on the detection performance. Section \ref{sec:imp&eva} reports the details of our implementation, the corresponding system overheads and the overall performance evaluation in an operational system on an indoor testbed. Finally, Section \ref{sec:conclusion} concludes the paper.

\section{Related Work}\label{sec:related}

Anomalous link quality degradation is 
a major cause of high packet losses in WSNs, reported by previous deployments \cite{6850017, 4408504, 6661323, 1182885}. Among various link quality metrics \cite{2240123}, RSSI provides direct channel quality information at the receiver, which is typically a required input for remedy systems in order to tune stack parameters such as transmission power \cite{Lin:atpc} or other layer parameters \cite{2185730,7164923}. Many studies \cite{levis2006rssi, 7164923} analyzed the relation between RSSI and packet loss and studied the temporal properties of RSSI \cite{2893729}. Only very few works have investigated how to use RSSI readings to detect good links (with low packet losses) turning into weak links (with high packet losses) with \textit{optimal} performance, i.e., \textit{robust detection with minimal detection error}. Our work is related to previous research on two topics: (1) network monitoring approaches for detecting link-related failures, and (2) anomaly detection techniques developed for WSNs. 

Existing approaches of network monitoring and diagnosis generally rely on active collection of node and network status. Some of them are centralized approaches, e.g., Sympathy \cite{1367278} and Emstar \cite{1267061}, in which a large amount of status information from individual sensor nodes (e.g., packet counter) is delivered to the sink to determine the failure causes. Agnostic Diagnosis \cite{6661323} constructs correlation graphs at the back-end server from collected system metrics to detect link failures. Other approaches, e.g., self-diagnosis \cite{6850017}, avoid sending all information to the sink by encouraging multiple sensor nodes to exchange information for cooperative failure detection. All these systems are powerful at detecting various failure types. However, they introduce large communication overhead to energy-constrained WSNs. Besides, \textit{most of these systems use metrics like packet counter or retransmission counter}. Though such metrics enable easy detection of packet losses, they can hardly be used to determine the cause, e.g., whether a loss is due to bad channel condition or packet collision. Instead, RADIUS utilizes RSSI, a channel quality attribute resident within every received packet, which does not require active information collection, allowing fully distributed anomaly detection. 

The research on anomaly detection in wired and wireless ad hoc networks is quite mature, but only a few solutions can be directly applied to WSNs due to the limited memory and computational capability of sensor nodes. Data mining and computational intelligence-based techniques, such as clustering \cite{4085803}, support vector machine \cite{4289308} and neural networks \cite{4797294}, own strong detection generality and accuracy as long as adequate attributes are in use \cite{1988328}. However, they all come with high complexity. In addition, they often rely on a central entity to cope with heavy tasks. PAD \cite{5356174} deduces link level errors with a probabilistic inference model maintained at a server. Statistical techniques such as kernel density estimator also require high computational capability to generate the density estimator. A recent work of employing such techniques is RASID \cite{6199865}, implementing the system on more powerful devices (WiFi access points) to detect intruders. Due to the limited resources of sensor nodes, data mining or machine learning oriented approaches are normally infeasible for distributed anomaly detection systems in WSNs.

The most widely used anomaly detection technique in WSNs is the statistical measure-based technique (e.g. mean, variance, maximum, self-defined) due to its low complexity and high effectiveness of finding detection boundaries (i.e. thresholds). For example, Fine-grained Analysis \cite{2517408} detects security attacks when RSSI changes exceed the measured maximum fluctuation occurred during the initial training phase. In the statistical measure category, there are two often used techniques: (1) \textit{CDF-based} thresholding (or percentile-based thresholding), and (2) \textit{Chebyshev inequality}-based thresholding. In CDF-based schemes, the threshold is defined as the \textit{x-th percentile} of the underlying data distribution of the monitored attribute. An example is the Memento system \cite{4068315} where an empirical CDF of consecutively missing heartbeat numbers is used to detect sensor failures. Another example is RASID \cite{6199865} which also defines a threshold at a given percentile after the density function is estimated. In other cases, when the underlying probability distribution of the monitored attribute is not known \textit{a priori}, the Chebyshev thresholding technique has often been applied. For instance, Chebyshev thresholding is used in \cite{1592596} to troubleshoot the network performance issues. 
In \cite{1689248}, a fusion threshold bound is derived using the Chebyshev inequality for target detection in WSNs. 

Despite their low complexity and easy adaptation to WSNs, both CDF-based and Chebyshev inequality-based methods are not designed for optimizing detection accuracy. In addition, achieving robust performance in detection accuracy by them is also a challenge. Later in this paper, we show that employing these two methods to achieve best detection accuracy implicitly requires manual fine-tuning of the threshold parameters
for each monitored link, which is difficult to do in practice. In RADIUS, we employ the Bayes decision theory \cite{melsa1978decision} to minimize the detection error, which is also a thresholding technique and has been widely used in other fields, e.g., signal detection and image analysis. Specifically, we apply the Bayesian thresholding technique to identify good links and weak links based on the monitored RSSI values. Its complexity is similar to that of the CDF or Chebyshev thresholding technique. Additionally, we combine the Bayesian thresholding technique with several supporting techniques to build a robust and accurate system for detecting anomalous link quality degradation in WSNs. 

\section{The RADIUS System Design}\label{sec:system}
In this section, we give an overview of the system design and the architecture of RADIUS, followed by the introduction of its major functional modules. In addition, we motivate the use of Bayesian thresholding in RADIUS to achieve minimal error rate and high robustness in detecting link quality anomalies, based on a comparison with the CDF thresholding and the Chebyshev thresholding techniques.  


\subsection{RADIUS System Overview}
\begin{figure}[t]
	\centering
	\includegraphics[width=1\linewidth]{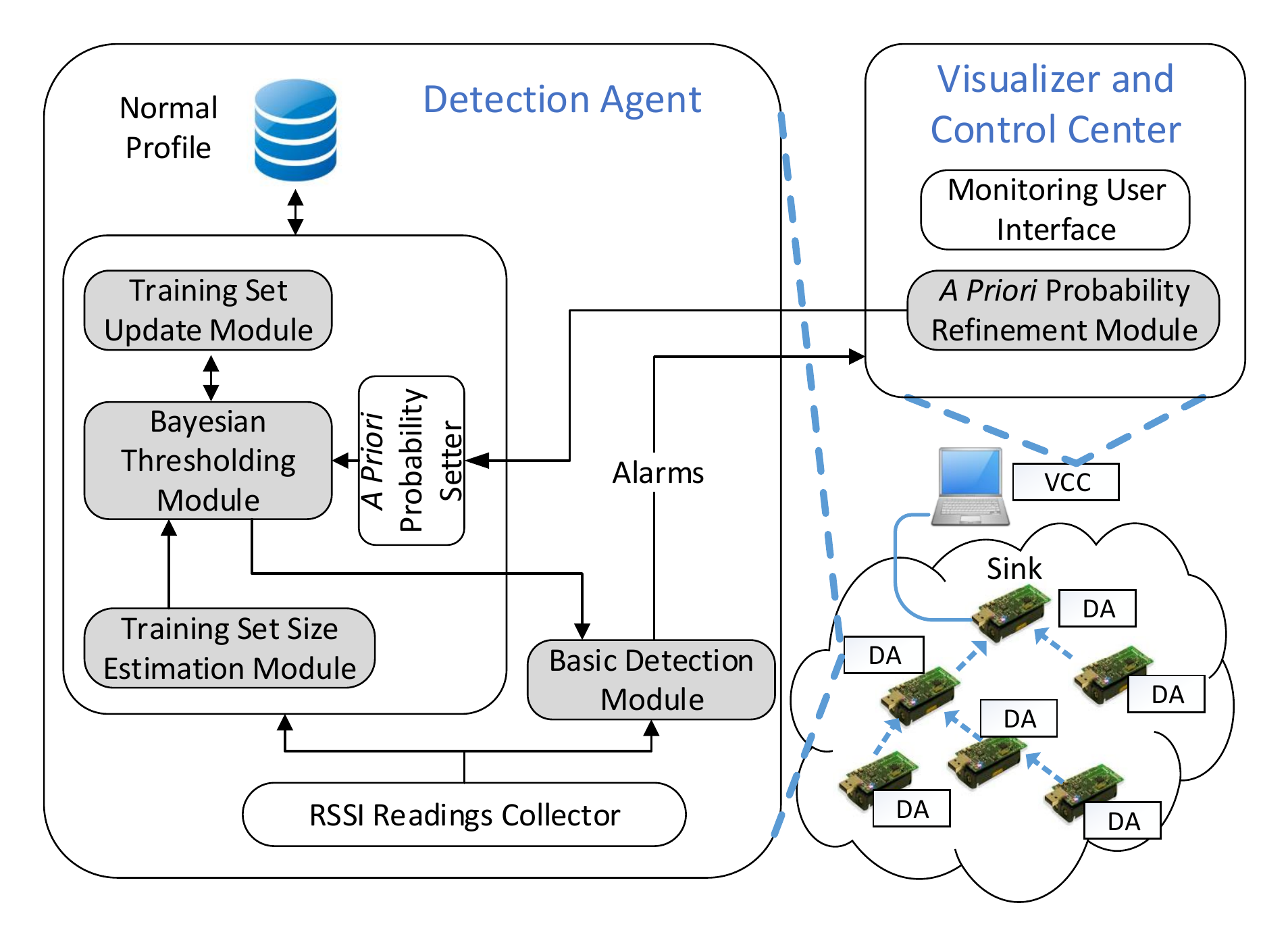}
	\vspace{-0.9cm}
	\caption{\textbf{The RADIUS system design and its functional modules.}}
	\label{fig:architecture}
	\vspace{-0.8cm}
\end{figure}

To achieve its goal, RADIUS adopts a hybrid approach. It comprises a set of distributed software modules located at the sensor nodes, which are called \textit{Detection Agents} (DAs), to detect anomalous link quality degradation along routing paths. In addition, a central server, called \textit{Visualizer and Control Center} (VCC), is used to monitor the performance of both the network and the RADIUS system. The overall system architecture of RADIUS is shown in Figure~\ref{fig:architecture}.

\paragraph{Two phases.} Similar to other anomaly detection systems, RADIUS runs in two phases: a training phase and an anomaly detection phase. During the \emph{training phase}, the user first observes if the performance of the network, e.g., packet delivery rate, is above the user requirement. In this case, each DA measures and collects the RSSI readings of the received packets to construct a ``normal profile'' for each monitored link, based on which a set of thresholds are generated. In the following \emph{anomaly detection phase}, each DA compares the runtime RSSI readings with the generated thresholds to detect if there is an anomalous link quality degradation.


\paragraph{System modules.} Choosing an appropriate anomaly detection technique is the key to achieve high detection performance, especially when using the highly varying RSSI values to distinguish the good links (with low packet losses) from weak links (with high packet losses) as accurately as possible. To tackle this problem,  RADIUS employs a thresholding technique based on the Bayes decision theory. 
Through the process of deciding a Bayes threshold, the detection error rate
can be minimized, as to be discussed in Section \ref{sec:thdComputation}. In a (close to) Gaussian channel, the computation of the Bayes threshold for a desired error only relies on a user-defined parameter (i.e. the \textit{a priori} probability in the Bayes formula) and two statistical measures (i.e. the {\em mean} and the {\em standard deviation}) of the measured RSSI values. 

During the training phase, the \texttt{Bayesian Thresholding Module} in each DA constructs the RSSI profiles (mean and standard deviation) of good links, and at the end of this phase,  the module derives a Bayes threshold for each monitored link. Note that, although the detection error is minimized by employing the Bayesian thresholding technique, the detection error rate may still be high when the mean and standard deviation estimates are not accurate due to insufficient RSSI training samples. To alleviate this problem, RADIUS employs a \texttt{Training Set Size Estimation Module} before the threshold computation. In this module, we use a confidence interval to estimate for each link the number of minimal RSSI samples required to produce a good approximation of the true underlying distribution of the RSSI values and hence the mean and standard deviation. This module ensures a ``good enough'' quality of the first Bayes threshold while keeping an acceptable training set size to avoid an overly long training time.

After the above operations, the anomaly detection phase starts. Recall that RADIUS is aimed to achieve high detection accuracy for links under diverse link conditions and be robust to environment changes. This is realized mainly by the \texttt{Basic Detection Module} and the \texttt{Training Set Update Module}. For the former, i.e. to achieve consistent high detection accuracy for different links, the system should avoid fine-tuning of the parameters involved in the thresholding, which specifically means no tuning of the \textit{a priori} probability for each individual link. For this purpose, the \texttt{Basic Detection Module} compares the runtime RSSI measurements after smoothing with the Bayes threshold to decide about whether there is an anomalous link quality degradation or not. Later in Section~\ref{sec:ThresholdChoice}, we show that a near-optimal detection accuracy is possible for all links across the network by even a coarse choice of the \textit{a priori} probability setting. In addition, to cope with dynamic environment changes, the \texttt{Training Set Update Module} is introduced, which updates the RSSI training set continuously during the detection phase in a memory-efficient way.


To further explore the potential of the \textit{a priori} probability parameter in achieving or maintaining the high detection accuracy, a feedback-based \textit{a priori} probability adaptation technique is introduced in RADIUS during the anomaly detection phase through the \texttt{A Priori Probability Refinement Module}. If the module in the VCC observes that the error rate of RADIUS for a link increases above a certain threshold, it then informs the \textit{Setter} at the DA to tune the setting of the \textit{a priori} probability.

In summary, the DA on each sensor node monitors the link quality of the links used by the higher-layer network protocols and fire alarms to inform the VCC and the network administrator about detected anomalous link quality degradations. In the training phase, when the network performance satisfies the user requirements, each DA generates locally the best RSSI threshold using Bayesian thresholding (Section~\ref{sec:thdComputation}) for each monitored link after collecting enough samples as determined by the minimal training set size estimation (Section~\ref{sec:minTrainingSet}). In the detection phase, using the generated thresholds, each DA performs local detection including data smoothing (Section~\ref{sec:dataSmoothing}) and adaptively adjusts the threshold either with local information updating the training set (Section~\ref{sec:trainingSetUpdate}) or with the refinement of the \textit{a priori} probability by the feedback from the VCC (Section~\ref{sec:prioriRefinement}) to achieve high accuracy and robustness.

\paragraph{Remarks.} In RADIUS, we have focused on using the RSSI link attribute to detect anomalous link quality degradation. However, its approach is not strictly limited to using RSSI, which may be extended to use other communication attributes as potential indicators of (possibly other types of) network performance anomalies. For instance, \textit{packet CRC error rate} could be observed for identifying packet collisions, \textit{packet overflow rate} for indicating queuing losses, and \textit{packet inter-arrival time} for determining node crashes. 

\subsection{Motivation of Bayesian Thresholding} \label{sec:motivationBayes}

Before describing the details of the Bayesian thresholding technique, we first motivate and illustrate its need. Its ability to deal with the challenge of achieving minimized detection error in a noisy channel is then evaluated in comparison with the CDF-based (or Percentile-based) and Chebyshev inequality-based thresholding techniques.

\subsubsection{Achieving Minimal Detection Error} \label{sec:problem}
	
\begin{figure}[t]
	\centering
	\includegraphics[width=1\linewidth]{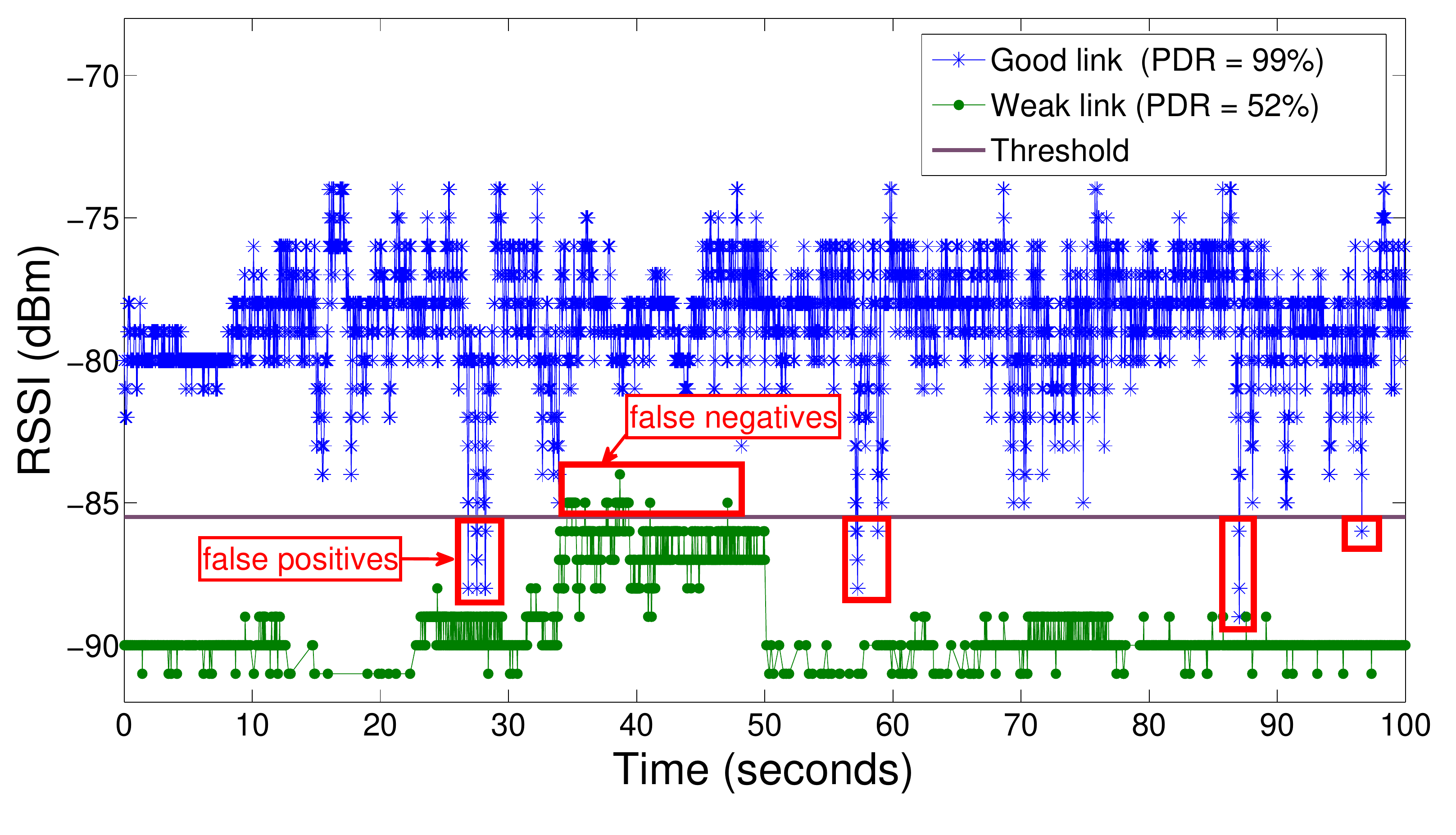}
	\vspace{-0.9cm}
	\caption{\textbf{The overlapping RSSI traces of a good link and a weak link implies that achieving high detection accuracy requires to minimize both the false positive and false negative rate.}}	
	\label{fig:PIMRC-RSSI}
	\vspace{-0.7cm}
\end{figure}

Due to its stochastic nature, the quality of a WSN link can vary randomly. To illustrate this, Figure \ref{fig:PIMRC-RSSI} presents two RSSI traces collected from a real WSN link. The upper one is when the link operated in a normal state with a packet delivery ratio (PDR) higher than 99\% (i.e. good link), while the lower one is when the link operated in an abnormal state with PDR below 52\% (i.e. weak link). The figure also shows that the two clusters of RSSI values, although mostly centred around their respective means, partially overlap with each other and no threshold can clearly discriminate them. This is due to the stochastic nature and the well-known temporal properties of low-power wireless links \cite{2893729}.

An implication of Figure \ref{fig:PIMRC-RSSI} is that no single RSSI threshold can, based on an RSSI value, lead to a definite conclusion without error if the link is in the good or weak state. Our objective is to find a threshold that minimizes the misidentification error. As shown in the figure, finding such a threshold involves a tradeoff decision to balance between reducing the chance that a good link is misidentified as a weak link and reducing the chance that a weak link is falsely viewed as a good link. The former is called false positive rate (FPR) and the latter called false negative rate (FNR). 

Mathematically, such a decision problem of finding the best threshold minimizing the error rate has been comprehensively studied under the Bayesian decision theory. In addition, if the monitored attribute is a Gaussian random variable (e.g., RSSI in a Gaussian channel), the complexity of Bayesian thresholding decreases significantly, falling within the limited capability of sensor nodes. This motivates our adoption of Bayesian thresholding in RADIUS. 

\begin{figure*}[t]
	\centering
	\includegraphics[width=1.0\linewidth, height = 4.5cm]{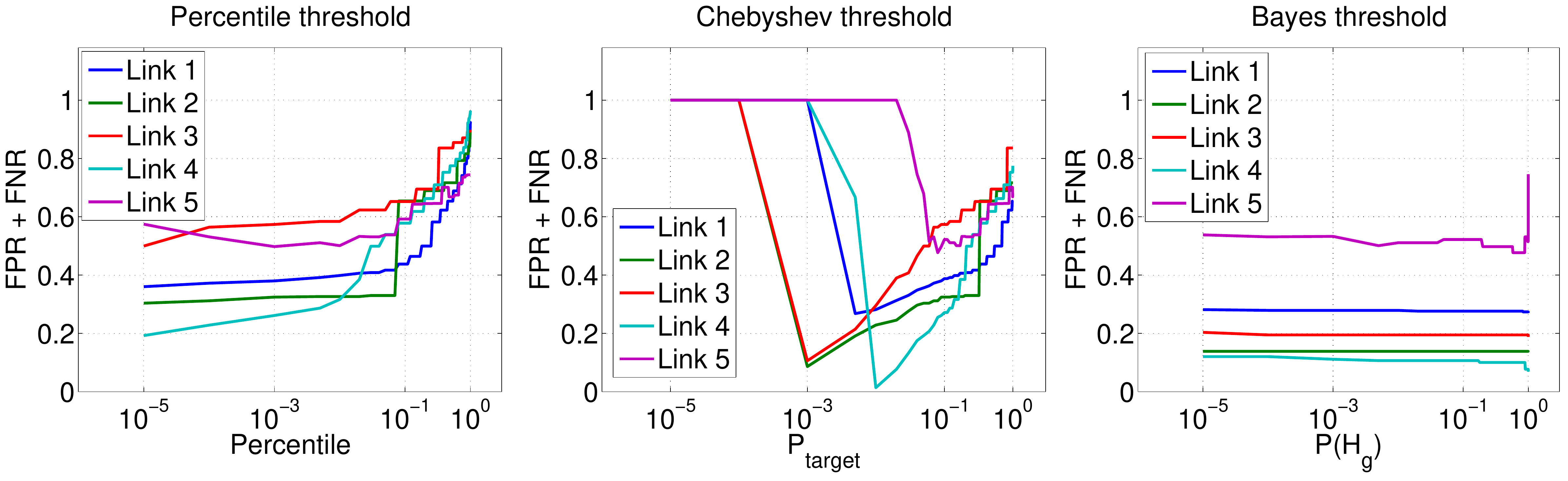}
	\vspace{-0.7cm}
	\caption{\textbf{The performance of CDF (Percentile) and Chebyshev thresholding varies dramatically depending on the parameter setting. Instead, the performance of Bayesian thresholding is robust, providing near-optimal accuracy under different link conditions with a coarse $P(H_g)$ setting.}}
	\label{fig:EVA-ThdCompare}
	\vspace{-0.55cm}
\end{figure*}

\subsubsection{Comparison among Thresholding Techniques}

To illustrate the high accuracy and robustness achieved by a Bayesian thresholding based detection system,  we compare the detection performance of Bayesian thresholding with that of two aforementioned popular statistical detection techniques in WSNs: CDF-based and Chebyshev-inequality based thresholding techniques, both of which have similar complexity as the Bayesian thresholding technique.  

In a CDF-based (or Percentile-based) detection scheme, the threshold is typically defined as the \textit{x-th} percentile of the underlying data distribution of the monitored attribute. For a Gaussian channel, the resulting threshold depends on the mean and standard deviation of the collected RSSI samples as well as on a parameter setting of the \textit{percentile} defined by the user.  The other thresholding technique often used in the literature of WSNs is the Chebyshev-inequality based technique.
In this case, the threshold is defined as follows:
\setlength{\belowdisplayskip}{3pt} \setlength{\belowdisplayshortskip}{3pt}
\setlength{\abovedisplayskip}{3pt} \setlength{\abovedisplayshortskip}{3pt}
\begin{equation}\label{equ:chebyTHD}
T_{cheby} = \overline{m} +  \sigma_m \ast \sqrt{\frac{1-P_{target}}{P_{target}}},
\end{equation}
where, in addition to the mean ($\overline{m}$) and standard deviation ($\sigma_m$) of the monitored attribute $m$, $P_{target}$ is a user-defined parameter for the desired false positive rate. Similarly, the Bayes threshold for Gaussian random RSSI depends on the same statistical measures (mean and standard deviation) as the CDF and Chebyshev methods and hence incurs a similarly low computation and memory overhead. The main difference is the parameter involved in the computation of the Bayes threshold: the \textit{a priori} probability of a link being in a good state ($P(H_g)$, as to be discussed in Section \ref{sec:thdComputation}).


For this comparison, we evaluate the system performance in terms of accuracy and robustness, with a focus on how the performance is influenced by the user-defined parameters, namely the \textit{x-th percentile} for CDF thresholding, $P_{target}$ for Chebyshev thresholding and $P(H_g)$ for Bayesian thresholding. The preferable technique is the one whose best detection accuracy performance is least sensitive to its parameter choice. Ideally, such a technique does not require fine-tuning of its parameter to achieve consistently optimal accuracy. 

We apply these three techniques individually to the same data traces collected from 5 different links in a network. The links are selected in such a way that diverse link conditions (e.g., line-of-sight,  non-line-of-sight, no human movements or frequent movement etc.) are captured. We record the false positive rate (FPR) and false negative rate (FNR) for each technique. A detection decision is considered as false positive (false negative) when the technique declares that an RSSI anomaly is detected (not detected) while the packet delivery rate over the link is above (under) a minimum (e.g., 80\%). The values of the above parameters of each method are varied in the same wide range of [$10^{-5}$, $1-10^{-5}$]. The resultant overall detection error rate (sum of FPR and FNR) for the three methods are presented in Figure \ref{fig:EVA-ThdCompare}.
	
From Figure \ref{fig:EVA-ThdCompare}, we can clearly see that the system performance with the Bayes threshold is consistently \textit{robust} for different links and is \textit{insensitive} to the value of $P(H_g)$ unless $P(H_g)$ is approaching the extreme, i.e., 1. 
In contrast, the system performance of both Percentile-based and Chebyshev-based approaches dramatically varies with changing parameter values. Furthermore, the optimal $P_{target}$ for Chebyshev threshold with minimal detection error varies significantly from link to link. Instead, Bayesian thresholding with a global coarse setting of $P(H_g)$ (e.g., any value from 0.1 to 0.9) for the analyzed 5 links provides close to the minimal detection error achieved by the Chebyshev thresholds. Finally, the accuracy with the Percentile threshold is in general worse than that with the Bayes threshold for every link with any parameter setting. 

An implication is that while CDF thresholding tends to provide tight RSSI bounds of good links for achieving a low FNR, it can easily cause a significantly high FPR in case of high randomness and temporal variations, making the technique difficult to achieve a low error rate. In addition, while Chebyshev thresholding can reach high detection accuracy, its performance highly depends on the choice of $P_{target}$. Bayesian thresholding, on the other hand, is designed to minimize the detection error while at the same time it can use a single setting of the threshold parameter $P(H_g)$ for the entire network, thus avoiding parameter tuning for every individual link. Thanks to such features, we employ Bayesian thresholding as the core detection technique in RADIUS. Details of Bayesian thresholding and the supporting techniques needed to use it are described in the following section.

\section{The RADIUS Techniques}\label{sec:approach} 

RADIUS aims to achieve robust and accurate detection for maintaining the network performance by combining the Bayesian thresholding with several supporting techniques. 
In this section, we present the Bayesian thresholding technique, elaborate the supporting techniques and identify the involved parameters. 


\subsection{Bayesian Thresholding} \label{sec:thdComputation}

A classical example that employs Bayesian thresholding is the binary detection problem, as known in the communication literature \cite{lee2012digital}. The goal is to detect binary digits ``0'' and ``1'' in a noisy channel based on the received signal level with the knowledge of the \textit{a priori} probabilities of ``0'' and ``1''. Mapping such a problem to our problem of detecting link quality degradation, our goal is to detect a link either being a good link (with low packet losses) or a weak link (with high packet losses) with a minimized error rate based on the RSSI values measured at the receiver of the link.

\paragraph{Mathematic Basis.} Let $H_g$ and $H_w$ respectively denote a link being a good and a weak link. Let $E$ denote a detection error (either a false positive or a false negative). Then, based on the Bayes decision theory, $P(E)$, the probability of detection error, can be expressed in terms of conditional probabilities as follows:
\setlength{\belowdisplayskip}{4pt} \setlength{\belowdisplayshortskip}{4pt}
\setlength{\abovedisplayskip}{4pt} \setlength{\abovedisplayshortskip}{4pt}
\begin{equation} \label{equ:bayesRule}
	P(E) = P(E|H_g)P(H_g) + P(E|H_w)P(H_w),
\end{equation}
where $P(H_g)$ is the \textit{a priori} probability of a link being a good link, and $P(H_w)=1-P(H_g)$. $P(E|H_g)$ is the probability of false positives, i.e., misidentifying a good link as a weak link, while $P(E|H_w)$ is the probability of false negatives, i.e., failing to detect a degradation in link quality. 

\begin{figure}
	\centering
	\includegraphics[width=1\linewidth]{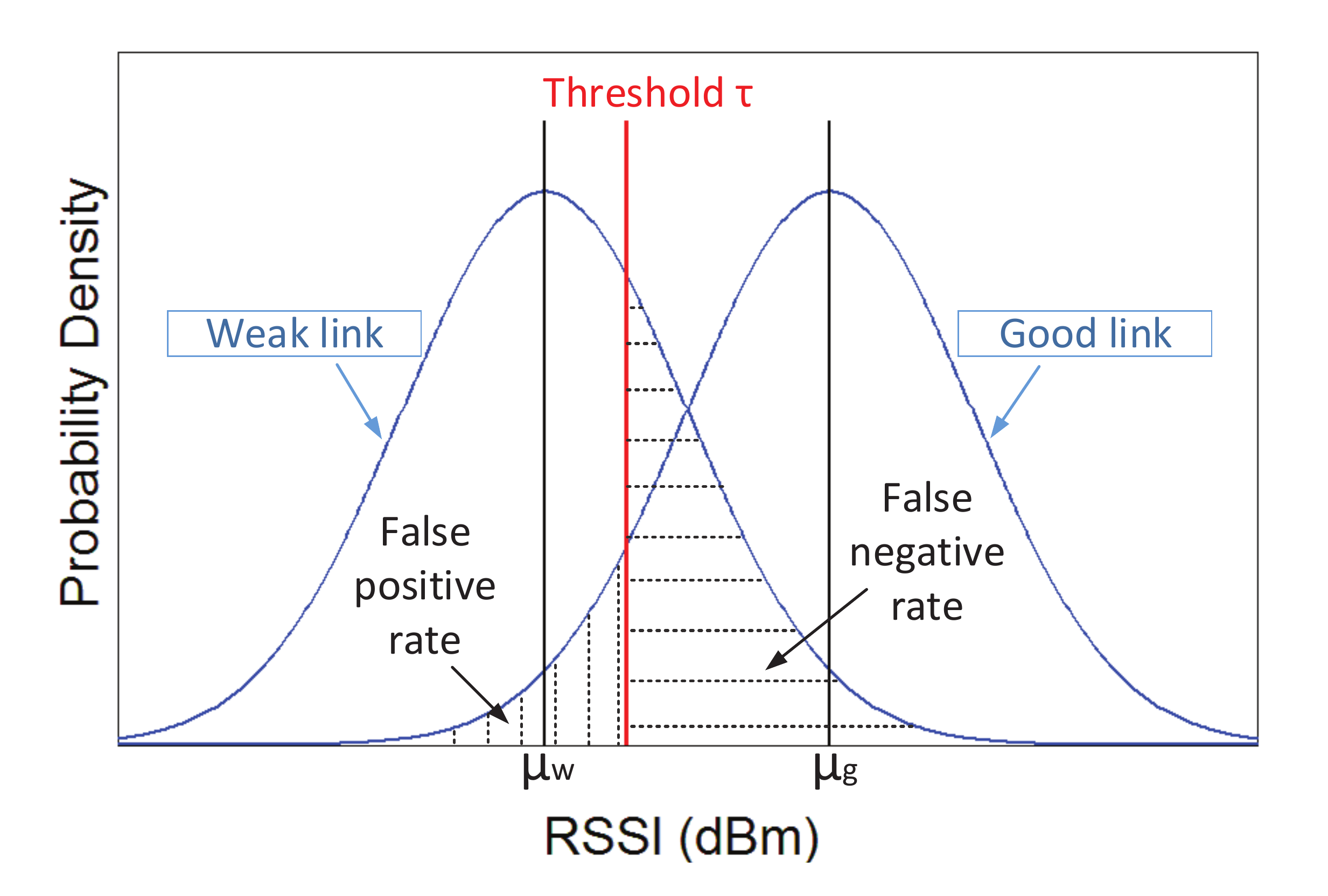}
	\vspace{-0.8cm}
	\caption{\textbf{Illustration of the probabilities of error for a binary classification problem with Gaussian distribution.}}
	\label{fig:rssi-threshold}
	\vspace{-0.8cm}
\end{figure}

We assume that the RSSI follows a normal distribution $N(\mu, \sigma)$,
which has been experimentally validated for low power communication in WSNs \cite{7164923, Rappaport:1996:WCP:525688}. For simplicity, we further assume that the distributions of RSSI for a weak link and for a good link, while with different means  $\mu_w$ and $\mu_g$ respectively, have the same standard deviation $\sigma$. In other words, the probability density functions of RSSI for the good link $f_g(x)$ and for the weak link $f_w(x)$ are as follows:
\begin{eqnarray}
	f_g(x) &=& \frac{1}{\sqrt{2\pi}\sigma}exp\left\lbrace -(x-\mu_g)^2 / 2\sigma^2\right\rbrace \label{equ:pdfWeak} \\
	f_w(x) &=& \frac{1}{\sqrt{2\pi}\sigma}exp\left\lbrace -(x-\mu_w)^2 / 2\sigma^2\right\rbrace \label{equ:pdfGood}
\end{eqnarray}
An example of these density functions is plotted in Figure \ref{fig:rssi-threshold}, where the false positive rate and the false negative rate are marked as shaded areas with respect to an RSSI threshold $\tau$. 

Based on Equations \ref{equ:pdfWeak} and \ref{equ:pdfGood}, the Bayes error $P(E)$ can be expressed as a function of the threshold $\tau$:
\begin{equation} \label{equ:fpr}
	P_E(\tau) = \int_{-\infty}^{\tau} f_g(x) dx \cdot P(H_g) + \int_{\tau}^{\infty} f_w(x) dx \cdot P(H_w)
\end{equation}


We minimize $P(E)$ by letting $d\Big(P_E(\tau)\Big)/d\tau = 0$. We can then obtain the optimal threshold $T_{Bayes}$ that minimizes the detection error and the resultant Bayes error $P_E$ as follows:
\begin{align}  
	T_{Bayes} &= \frac{1}{2}(\mu_g + \mu_w) + \frac{\sigma^2 ln(P(H_w)/P(H_g))}{\mu_g - \mu_w} \label{equ:rssiTHD} \\
	P_E(\alpha) &= Q\bigg[\alpha - \frac{1}{2} \alpha^{-1} ln\Big(P(H_w)/P(H_g)\Big)\bigg]P(H_g) \nonumber \\
	& + Q\bigg[\alpha + \frac{1}{2} \alpha^{-1} ln\Big(P(H_w)/P(H_g)\Big)\bigg]P(H_w) \label{equ:error}   
\end{align} 
where $Q(x)$ is a Q-function \cite{wiki:qfunction} and $\alpha$ is defined as: 
\begin{equation} \label{equ:similarityMetric}
	\alpha = (\mu_g - \mu_w)/2\sigma. 
\end{equation}

\paragraph{Application to RADIUS.} Applying the Bayesian thresholding technique to RADIUS essentially requires to find the Bayes threshold $T_{Bayes}$, as computed in Equation \ref{equ:rssiTHD}. The use of Equation \ref{equ:rssiTHD} also indicates that the computation of the Bayes threshold only depends on a few statistical measures, significantly reducing the complexity in comparison to typical Bayesian decision problems. These statistical measures include the mean of the density distribution of RSSI for a good link and for a weak link ($\mu_{g}$ and $\mu_{w}$, respectively), as well as the standard deviation $\sigma$ of the distribution. 

In the RADIUS system, the VCC monitors the end-to-end packet delivery ratio (PDR) of the network after the network is deployed. When the PDR satisfies the minimal user requirements, implying that all links being used are good links, the VCC informs all DAs to start the training phase. During this phase, the statistical measures $\mu_{g}$ and $\sigma$ are computed from the collected RSSI samples for each individual link relevant to the higher-layer routing protocols. For the value of $\mu_{w}$, we choose the border RSSI value of the ``grey zone'' (i.e., $\mu_w = -88$ dBm) reported in previous experimental studies \cite{Lin:atpc, 7164923}, which show that PDR decreases significantly when a link enters the ``grey zone''. 

The \textit{a priori} probability $P(H_g)$ required in Equation \ref{equ:rssiTHD} is typically computed empirically based on previous experience or measurements. As this parameter has to be defined before the deployment, the setting of $P(H_g)$ impacts on the detection performance of the operational system. In Section \ref{sec:motivationBayes}, we have presented its effect on the detection error in general. In Section \ref{sec:ThresholdChoice}, we will quantify the effect of $P(H_g)$ on the FPR and the FNR in more details.

\paragraph{Bayesian thresholding alone is not enough.} Despite that the Bayes threshold is designed in RADIUS to minimize detection error, it alone is not enough to build a robust and accurate system for the detection of anomalous link quality degradation in WSNs. There are several inherent challenges that have to be addressed in order to apply the thresholding technique. Some of them are introduced by the technique itself (e.g., accurate estimation of $\mu_{g}$ and $\sigma$) while the others are caused by the fact that RSSI is highly influenced by environment changes (e.g., smoothing and threshold adaptation). In the rest of this section, we introduce these challenges and the techniques that we employ in RADIUS to address them.

\subsection{Estimating the Minimal Training Set Size}\label{sec:minTrainingSet}

According to Equation \ref{equ:rssiTHD}, finding the Bayes threshold requires to compute the mean $\mu_{g}$ and the standard deviation $\sigma$ from a collected RSSI training set. The size of such a training set has a significant impact on the estimation error of $\mu_{g}$ and $\sigma$ and hence a great impact on the system detection accuracy. Deciding the training set size involves considering a tradeoff between the detection accuracy and the training latency. A larger training set can provide a more accurate estimation of the underlying data distribution thus higher accuracy in estimating $\mu_{g}$ and $\sigma$, while it may significantly increase the training time. In addition, the training size should also differ from link to link for the specific statistical characteristics of individual links: small training set size may suffice for a stable link while a larger size is required for links with highly fluctuating RSSI readings. 

To address this challenge, we analyze the confidence interval to estimate the minimal training set size of RSSI values for each individual link. In this way, RADIUS achieves a good tradeoff between system detection accuracy and training latency. After the training phase starts, the minimal training set size is decided by the DAs for each individual link after the collection of a few RSSI samples.

In particular, each DA first computes the standard deviation $\sigma_s$ for the first $N_s$ samples of RSSI collected in a short time period. Then, the DA estimates the minimal training set size $N_{ts}$ for a given error $E_{\mu}$. According to the \textit{Central Limit Theorem}, for an attribute $x$ with any type of underlying distribution, the margin error of the confidence interval for the attribute mean $\bar{x}$ is  $e_{\mu} = z\cdot\sigma_p/\sqrt{n}$, where $z$ is the z-score ($z=2.58$ for a confidence level of 99\%), $n$ is the number of samples and $\sigma_p$ is the population standard deviation. With this, the minimal training size $N_{ts}$ is calculated as:
\begin{equation} \label{equ:minSampleSize}
{N_{ts}} = {\left(\frac{z\cdot \sigma_p }{E_{\mu}}\right)}^{2}
\end{equation} 
where $E_{\mu}$ is a user-defined parameter for the maximum error of the estimated RSSI mean. In addition,  $\sigma_p$ can be substituted by the standard deviation $\sigma_s$ of the first $ N_s$ samples, which has to be larger than 30 \cite{pitman1993probability}. 

Applying Equation \ref{equ:minSampleSize} allows every DA to find appropriate RSSI training set sizes for each of its observed links, achieving a good tradeoff between estimation accuracy and training latency before computing the Bayes thresholds. Nevertheless, we need to find appropriate settings of $N_s$ and $E_{\mu}$ to apply Equation \ref{equ:minSampleSize}. In Section \ref{sec:evaParamTrainingSize}, we will show the impact of $N_s$ and $E_{\mu}$ on the training set size and subsequently on the overall detection accuracy of RADIUS and then suggest the parameter choices of $N_s$ and $E_{\mu}$ for an indoor environment.

\subsection{Data Smoothing} \label{sec:dataSmoothing}

As illustrated by Figure \ref{fig:PIMRC-RSSI}, the RSSI signal is random in nature. In other words, an RSSI value lower than the Bayes threshold may actually be attributed to its normal randomness while not due to anomalous channel quality degradation. As a consequence, comparing each individual RSSI value with the Bayes threshold $T_{Bayes}$ and using the comparison result to decide about an anomaly can lead to an over or under estimation of anomalies.  

To overcome this limitation and make RADIUS more robust, each DA applies a sliding window of size $l$ to compute a short-term average of RSSI and compares the $l$-averaged RSSI with the Bayes threshold to trigger an anomaly detection. 
Intuitively, the choice of $l$ has an influence on the detection accuracy. A smaller $l$ makes the detection more responsive, but it may not be sufficient to clean the noise. On the other hand, a larger $l$ may be a better choice for data cleaning, but overly smoothing may fail to capture abnormal events. To understand the impact of the sliding window size, we show the effect of $l$ on the system performance and suggest a proper setting for indoor environments in Section \ref{sec:ImpactDataSmoothing}.

\subsection{Updating the Training Set} \label{sec:trainingSetUpdate}

After the Bayes threshold is determined, RADIUS performs anomaly detection on the monitored RSSI values by comparing them against $T_{Bayes}$. While $T_{Bayes}$ is designed to minimize the detection error, the underlying RSSI distribution may vary due to environmental changes. Consequently, the mean and standard deviation estimated in the training phase may no longer be valid. This requires updating the estimated RSSI distribution, as well as $T_{Bayes}$ accordingly, in response to environmental changes.

To cope with dynamic changes in the environment, we update the mean and standard deviation of the RSSI distribution during the entire anomaly detection phase. The Bayes threshold is then updated using Equation \ref{equ:rssiTHD}. 
Specifically, we update the mean and standard deviation by updating the training set with the normal RSSI readings observed during the period of detection. To identify whether an RSSI value is normal or not, RADIUS assigns an anomaly score $a_{t}$ for it at time $t$, indicating the significance of the anomalous behavior. $a_{t}$ is calculated by $a_{t} = \bar{S}/T$, where $\bar{S}$ is the sliding window average of RSSI values and $T$ is the Bayes threshold. 

During the detection phase, RADIUS collects consecutive RSSI readings in disjoint groups of size $l_{update}$ and also their anomaly scores in a separate group to compute the average anomaly score. The group of RSSI readings with an average anomaly score of less than one is added to the training set. Through this, the training set is updated. This technique is similar to the silence profile updating scheme in \cite{6199865} for intrusion detection. RADIUS, however, needs to keep separate groups to store anomaly scores because the threshold may vary in the middle of an update process due to the \textit{a priori} probability refinement technique discussed in Section \ref{sec:prioriRefinement}, while it remains constant in the scheme used in \cite{6199865}.

To minimize the memory overhead due to the update process, we employ a memory-efficient way to update the RSSI mean and standard deviation without incrementing the buffer to store new RSSI samples. Details of its implementation and memory overhead are described in Section \ref{sec:overhead}. In addition, updating the training set requires a proper setting of the parameter $l_{update}$. In Section \ref{sec:evaParamTrainingUpdate}, we quantify the effect of $l_{update}$ on the performance of RADIUS and recommend the appropriate $l_{update}$ value for an indoor environment.

\subsection{Refinement of the A Priori Probability} \label{sec:prioriRefinement}

In addition to the RSSI mean and standard deviation that are measured and updated as previously discussed, Equation \ref{equ:rssiTHD} uses another parameter, $P(H_g)$. This parameter is an empirical, \textit{a priori} probability decided before system deployment. We showed earlier that the performance of RADIUS is not sensitive to the setting of $P(H_g)$. However, an initial coarse setting of $P(H_g)$, due to environment changes, may no longer provide the best detection accuracy and hence become outdated, requiring a refinement.

To address this challenge, RADIUS refines the \textit{a priori} probability $P(H_g)$ and then updates the Bayes threshold during the detection phase when necessary. Specifically, when an anomaly is detected and an alarm is triggered, the VCC keeps recording the number of false alarms. An alarm is considered as a false alarm if the PDR over the path of the anomalous link, when the alarm is received at the VCC, remains above a given threshold. RADIUS counts the number of consecutive false alarms. When the number exceeds a predefined value $N_{alarm}$, the VCC informs the specific DA and the DA adjusts $P(H_g)$, incrementing it by $\delta$ until reaching a predefined \textit{maximum} $P(H_g)$. Here, we call $\delta$ the adjustment step and \textit{maximum} $P(H_g)$ the allowed upper limit for the parameter to avoid over-adjustments that may lead to a significant increase in FNR. Each time when $P(H_g)$ is updated, the Bayes threshold is updated accordingly. 
In Section \ref{sec:ThresholdChoice}, we will analyze the effect of the initial and maximum setting of $P(H_g)$. In Section \ref{sec:EVA-paramRefine} we will discuss in details the effect of $N_{alarm}$ and $\delta$ on the performance of RADIUS and recommend the appropriate settings for an indoor environment.

\section{Setting the RADIUS Parameters}\label{sec:parameterChoice}

\begin{figure*}[t]
	\centering
	\includegraphics[width=1.0\linewidth, height = 5cm]{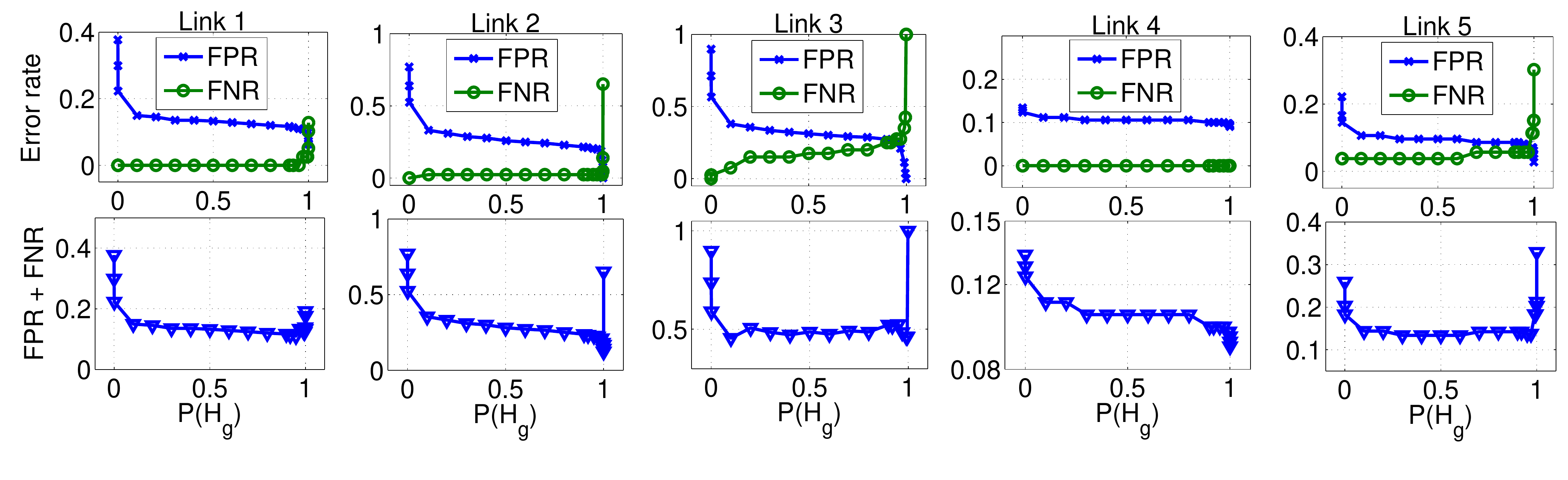}
	\vspace{-1.1cm}
	\caption{\textbf{Effect of the \textit{a priori} probability $P(H_g)$ on the error rates for 5 representative links. $P(H_g)$ varies in the range [$10^{-5}$, $1-10^{-5}$].} }
	\label{fig:EVA-ProbGood}
	\vspace{-0.4cm}
\end{figure*}

\begin{figure*}[t]
	\centering
	\begin{minipage}{.35\textwidth}
		\centering
		\includegraphics[width=1\linewidth, height=4cm]{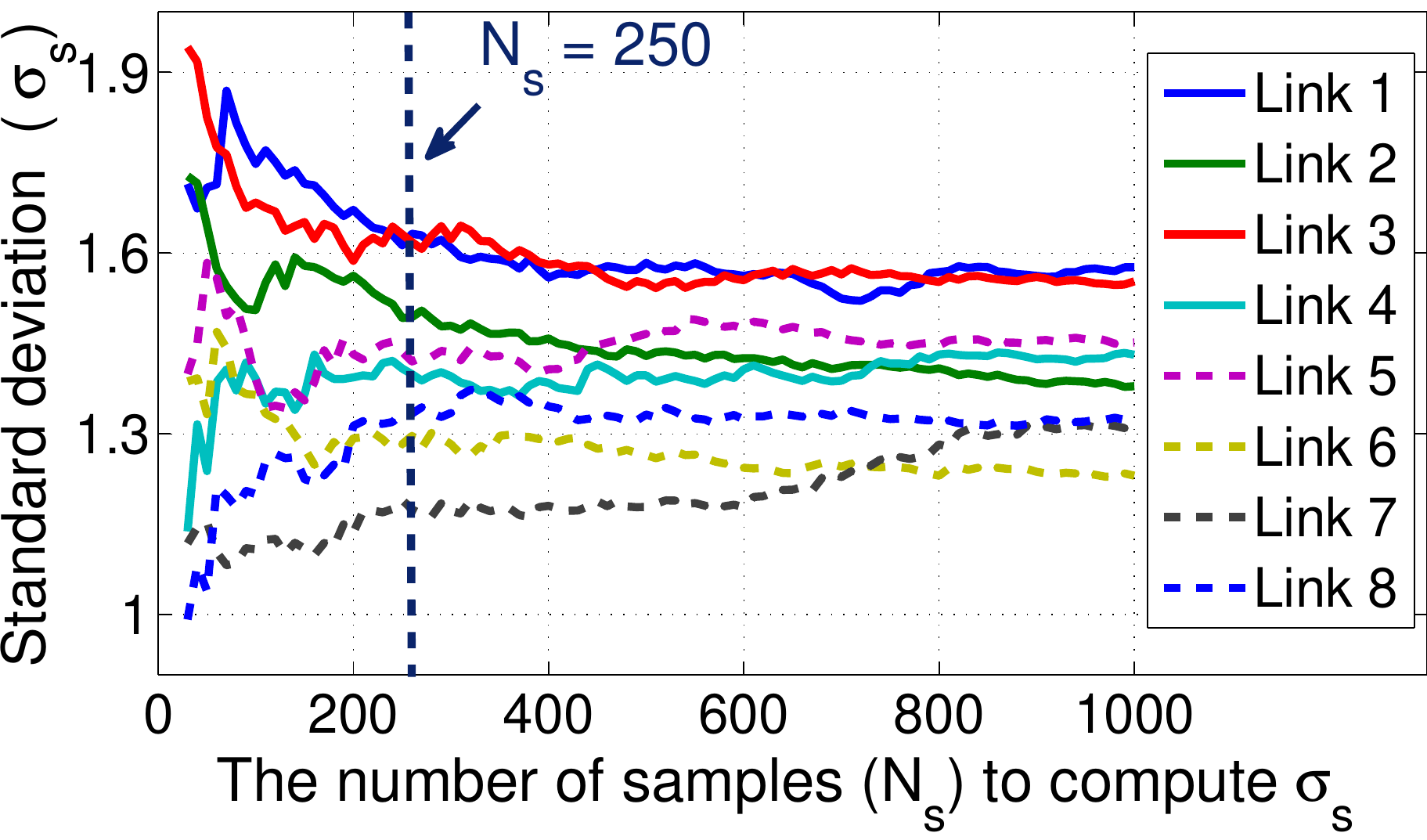}
		\vspace{-0.75cm}
		\captionof{figure}{\textbf{Effect of the parameter $N_s$. The resultant $\sigma_s$ becomes more stable after $N_s = 250$.} }
		\vspace{-0.5cm}
		\label{fig:Nsigma}
	\end{minipage} \hfill
	\begin{minipage}{.6\textwidth}
		\centering
		\includegraphics[width=1\linewidth, height=4cm]{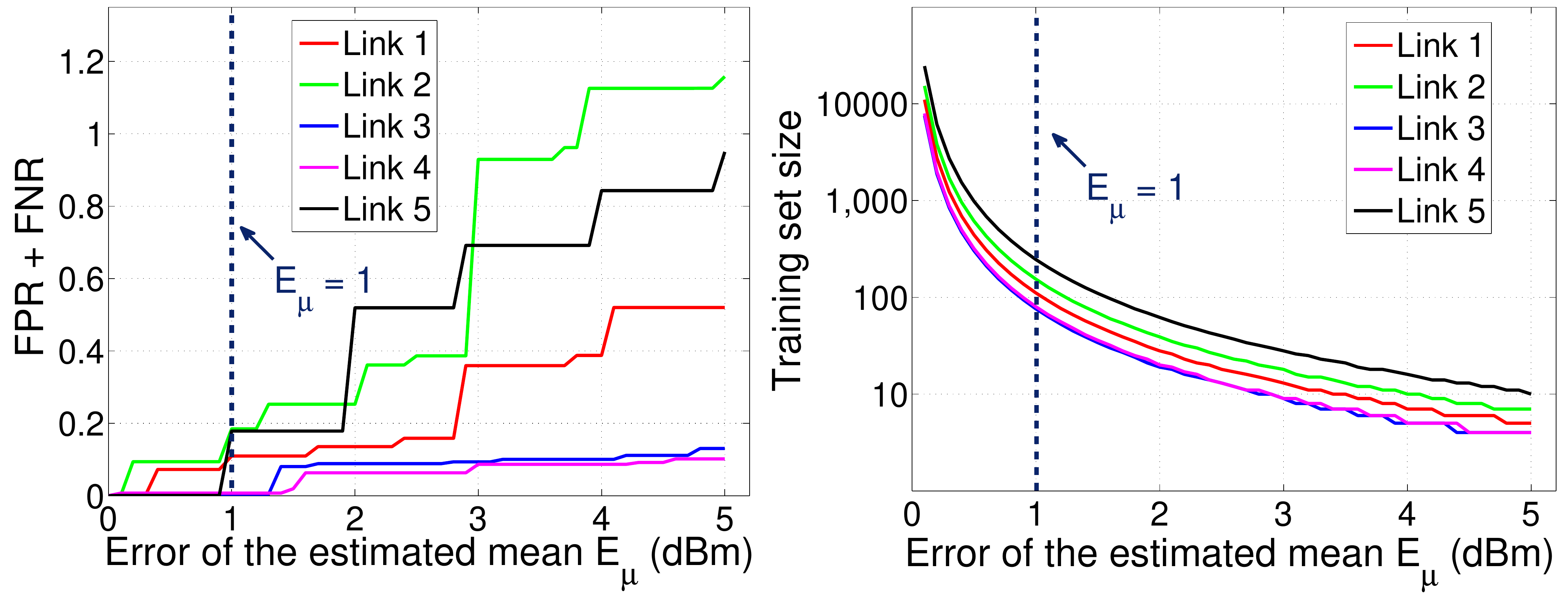}
		\vspace{-0.75cm}
		\captionof{figure}{\textbf{Effect of the parameter $E_{\mu}$. $E_{\mu} = 1$ dBm provides a good tradeoff between the detection accuracy and the training set size (i.e. training latency).}}
		\vspace{-0.5cm}
		\label{fig:ErrorMean}
	\end{minipage}
\end{figure*}

In the previous section, we presented the details of the Bayesian thresholding and the supporting techniques in RADIUS. We now study the impact of the aforementioned parameters required in each individual technique on the system performance. Specifically, we elaborate the effect of the Bayes threshold parameter $P(H_g)$ and then explore the parameter space of all the parameters involved in supporting techniques. Based on detailed analysis, we give insights on the best parameter setting for an indoor office environment.

For this, we performed extensive experiments, collecting real data traces from an indoor testbed, whose details are reported in Section \ref{sec:evaluation}. To capture different link conditions, we selected 8 sender-receiver pairs (either line-of-sight or non-line-of-sight) at different locations with various environment dynamics (e.g., human movements and obstacles). In each experiment, we simulated a good link turning into a weak link by decreasing the transmission power of the sender node from the maximum level gradually to the minimum with a packet sending rate of 5 Hz. The receiver node records the RSSI and PDR traces for more than 15 minutes. We repeated the experiment 10 times for each link. The minimum PDR that decides whether a link is a good link or a weak link is set to 80\% throughout the whole analysis.

\subsection{Bayesian Thresholding} \label{sec:ThresholdChoice}

According to Equation \ref{equ:rssiTHD}, calculating the Bayes threshold
requires a user-defined parameter: \textit{a priori} probability $P(H_g)$.
Different from the general analysis depicted in Section \ref{sec:motivationBayes}, we present here the detailed analysis of the impacts of $P(H_g)$ on the false positive rate (FPR), false negative rate (FNR) and the total error rate. 
 
Figure \ref{fig:EVA-ProbGood} shows the change of the error rates with varying $P(H_g)$ ranging in [$10^{-5}$, $1-10^{-5}$] for 5 representative links out of the 8 analyzed links. We observe that the FPR always decreases with $P(H_g)$, while the FNR increases with $P(H_g)$. This is because a larger $P(H_g)$ indicates a higher weight on the FPR in the computation of the Bayes error (see Equation \ref{equ:error}). Hence,  reducing FPR is more effective than reducing FNR to keep the Bayes error rate low for a larger $P(H_g)$.  

Moreover, we can see in Figure \ref{fig:EVA-ProbGood} that the overall error rate mostly stays low regardless of the values of $P(H_g)$, except for the cases when the value of $P(H_g)$ is extremely close to 0 or 1. The results confirm that the system performance with the Bayes threshold is insensitive to the setting of $P(H_g)$ as long as extreme values are not considered. The reason for this is that the Bayesian thresholding approach always tries to balance between FPR and FNR for any $P(H_g)$ setting. 

From the figure, we can further see that a global $P(H_g)$ setting from a wide range (any value not close to 0 or 1) may not be the best setting for each individual link. However, it can provide for all different links near optimal detection accuracy at the same time. In other words, with a coarse global setting of $P(H_g)$ for all DAs, the Bayesian thresholding ensures RADIUS to deliver near optimal accuracy for different links under diverse link conditions without the need of tuning for each of them. For our case, we select the initial setting of $P(H_g)$ at 0.8 because we assume that the probability of the links being good is generally higher than that of being weak, in our deployment environment. Additionally, to avoid the significant increase of FNR caused by the over-adjustment due to the \textit{a priori} probability refinement, we limit the \textit{maximum} $P(H_g)$ to 0.99 for our deployment environment.

\subsection{Estimating the Minimal Training Set Size}\label{sec:evaParamTrainingSize}

As discussed in Section \ref{sec:minTrainingSet}, the first task of a DA before generating a Bayes threshold is to estimate the minimal training set size $N_{ts}$ for each link. A proper $N_{ts}$ needs to achieve a good tradeoff between detection accuracy and training latency. To apply Equation \ref{equ:minSampleSize}, the computation of $N_{ts}$ requires two parameters: (1) the number of first $N_s$ samples of RSSI for computing the standard deviation $\sigma_s$, and (2) the maximum error of the estimated mean  $E_{\mu}$. While the first $N_{s}$ samples of RSSI only give a quick indication, $N_{ts}$ is the resultant minimal training set size, from which the DA estimates the RSSI mean and standard deviation for computing Bayes thresholds. $N_{ts}$ is usually larger or at least equal to $N_{s}$.   

We first study the impact of $N_{s}$. A small $N_{s}$ may result in a partial view of the complete channel variation, while overly large $N_{s}$ may only increase the training delay. To understand the impact of $N_{s}$, we plot in Figure \ref{fig:Nsigma} the resultant standard deviation $\sigma_s$ for various values of $N_s$ based on the RSSI traces of all 8 links. We observe that the values of $\sigma_s$ initially have a larger variation and become more stable when $N_s$ is close to 250. The reason for this is that a small set of samples is insufficient to capture the overall temporal variations of RSSI, especially in an indoor environment where multi-path fading and interference are ubiquitous. Based on the result, we choose $N_s = 250$ for our indoor environment. 
  
Then we focus on the impact of $E_{\mu}$. According to Equation \ref{equ:minSampleSize}, the choice of $E_{\mu}$ has a tradeoff: smaller $E_{\mu}$ indicates higher estimation accuracy of the RSSI mean and thus higher detection accuracy; a smaller $E_{\mu}$, however, may also increase the training set size significantly. By varying the estimated errors $E_{\mu}$, we plot the resultant training set size and error rates for 5 representative links in Figure \ref{fig:ErrorMean}. 
The figure shows that the total error rate decreases significantly with a smaller $E_{\mu}$ at the expense of a rapidly increasing training set size. To balance between the detection error and the training time, we choose $E_{\mu} = 1$ dBm for our indoor environment, which causes only a slight increase of the detection error compared to that of $E_{\mu} = 0$ dBm while at the same time keeping $N_{ts}$ within the scale of a few hundred samples, achieving a good tradeoff between the training latency (several minutes with a sending frequency of 5 Hz) and detection accuracy. 

\begin{figure}[t]
	\centering
	\includegraphics[width=1.0\linewidth, height = 4.5cm ]{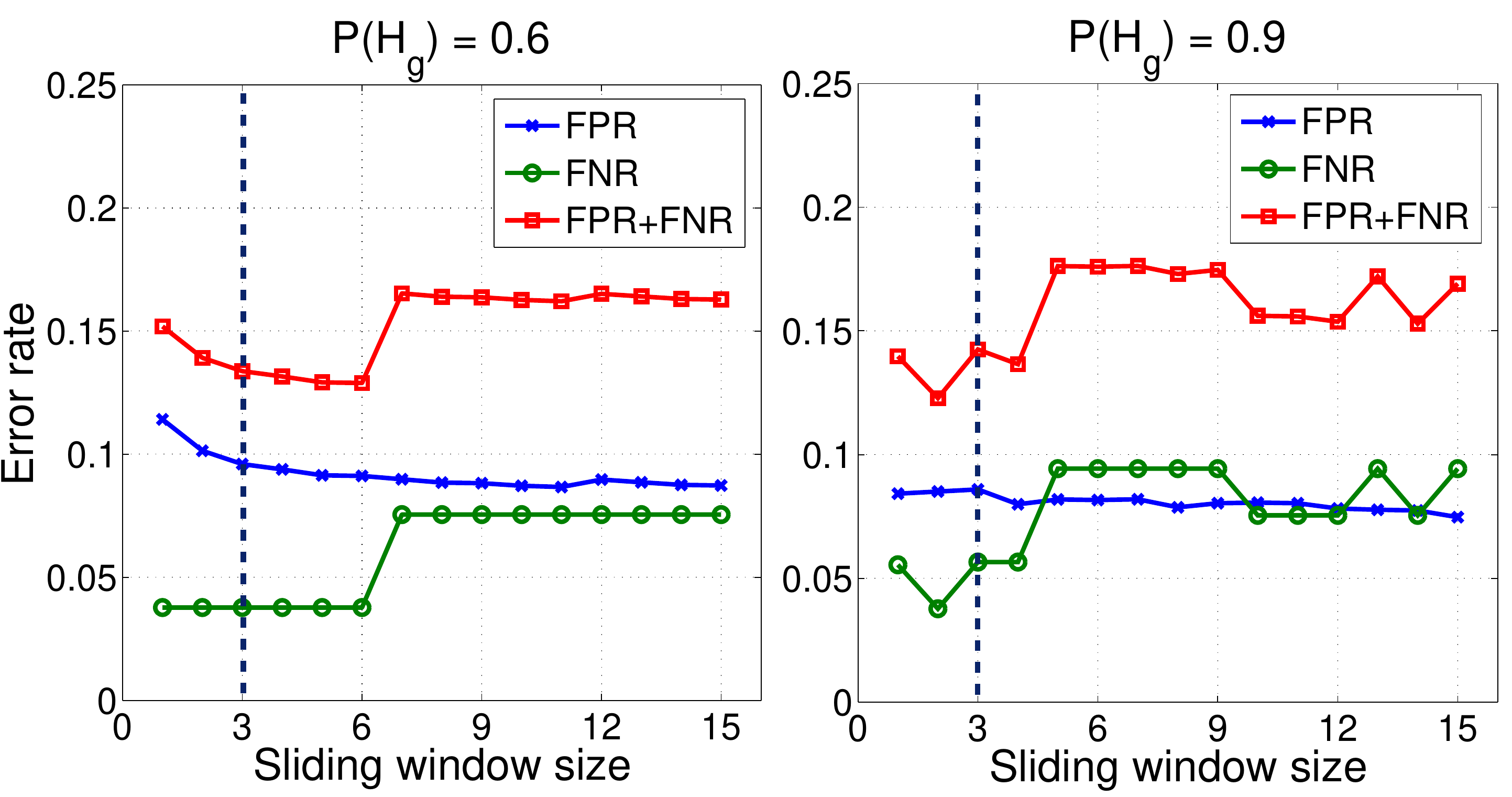}
	\vspace{-0.7cm}
	\caption{\textbf{Effect of data smoothing with different sliding window size $l$ for different $P(H_g)$. Error rate is close to lowest when $l=3$.} }
	\label{fig:EVA-slidingWindow}
	\vspace{-0.3cm}
\end{figure} 

\begin{figure}[t]
	\centering
	\includegraphics[width=1.0\linewidth, height = 6cm]{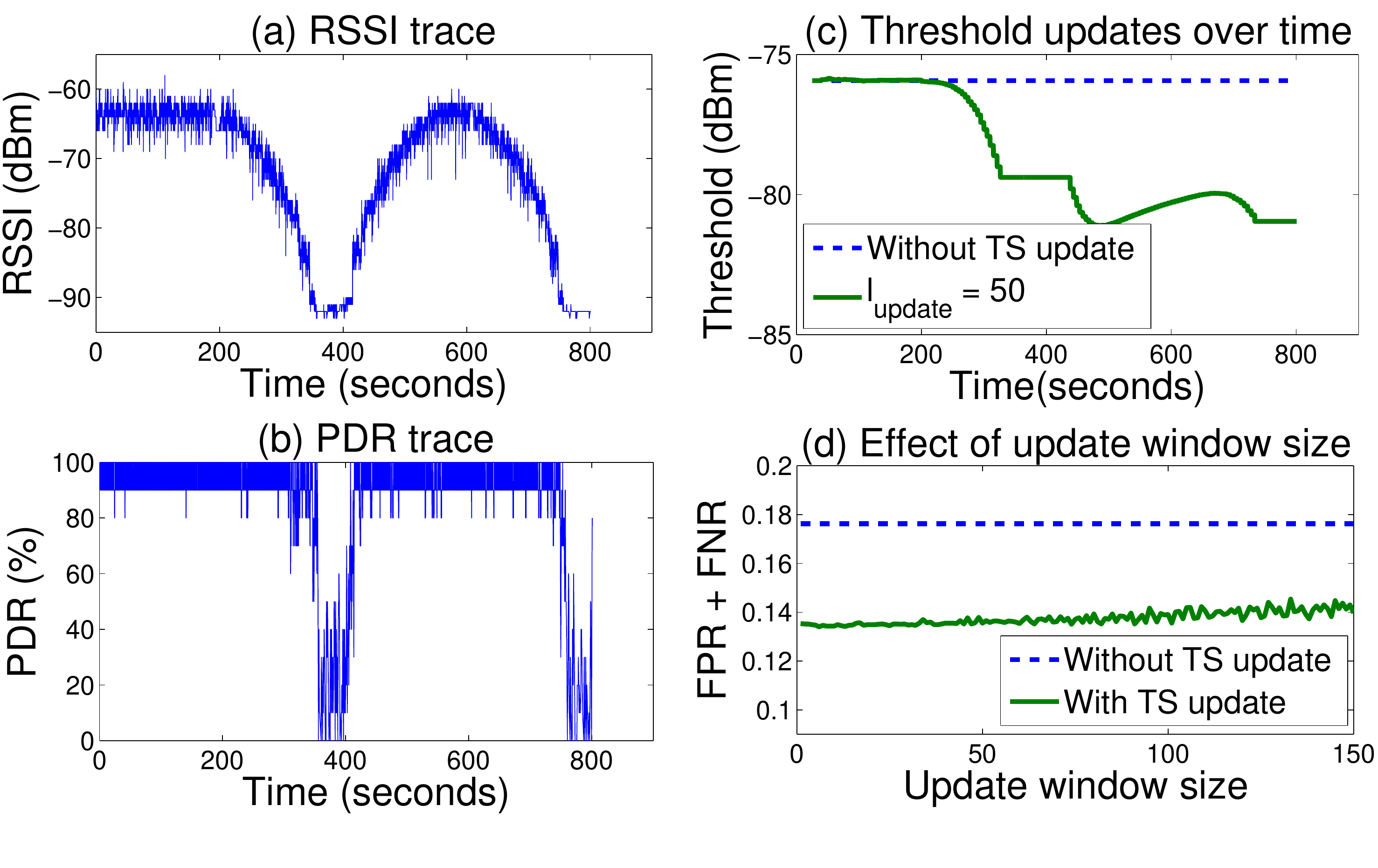}
	\vspace{-1cm}
	\caption{\textbf{Effect of training set (TS) update with varying update window size $l_{update}$.}}
	\label{fig:EVA-updateWindow}
	\vspace{-0.7cm}
\end{figure} 

\subsection{Data Smoothing} \label{sec:ImpactDataSmoothing}

As mentioned in Section \ref{sec:dataSmoothing}, smoothing the noisy data during the detection phase requires a sliding window of size $l$ to reduce the detection error caused by the normal RSSI randomness. To see the impact of $l$, we demonstrate how the error rate changes with different values of $l$ (window size from 1 to 15) under two representative $P(H_g)$ values.

We observe from Figure~\ref{fig:EVA-slidingWindow} that, for both $P(H_g)$ settings, increasing $l$ reduces FPR but increases FNR, which causes the total error rate to first decrease and then increase with a larger $l$. The reason is that smoothing RSSI is effective to reduce false alarms. However, if $l$ keeps increasing, at some point, the real RSSI anomaly events are smoothed out, causing a significant increase in FNR. The impact of $l$ is also related to the setting of the minimal PDR ($PDR_{min}$) that defines a good link. In our case, as PDR is computed over a sliding window of 10 packets and $PDR_{min}$ is set to 80\%, a small sliding window $l$ is preferred to avoid the significant increase in FNR. For our case, we choose $l = 3$, at which the total error rate is close to the lowest for both $P(H_g)$ settings.

\begin{figure}[t]
	\centering
	\includegraphics[width=1.0\linewidth, height = 8cm]{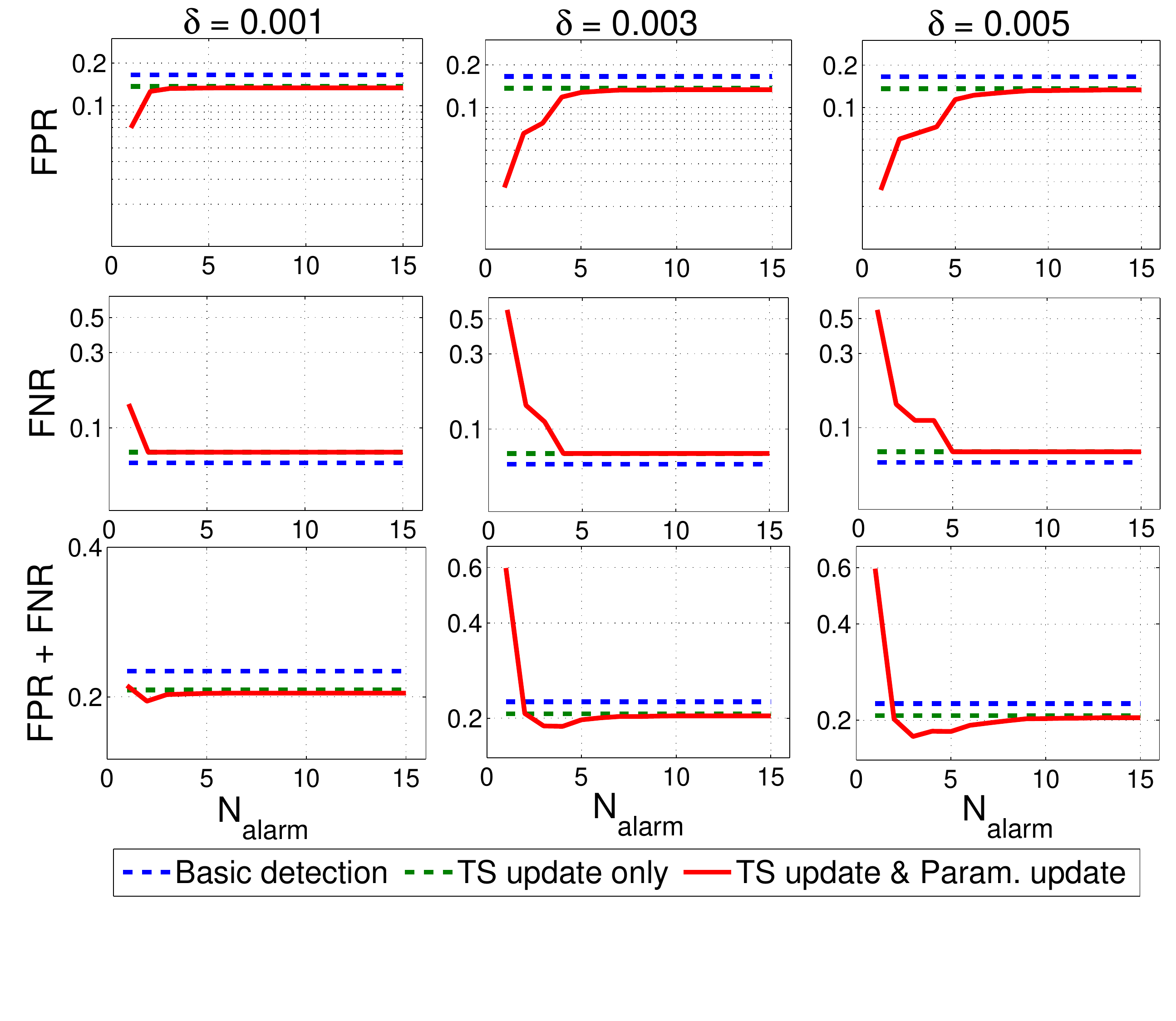}
	\vspace{-1.5cm}
	\caption{\textbf{Effect of the \textit{a priori} refinement with varying $N_{alarm}$ and $\delta$.}}
	\label{fig:EVA-paramRefine}
	\vspace{-0.65cm}
\end{figure}

\subsection{Updating the Training Set} \label{sec:evaParamTrainingUpdate}

To be adaptive to varying environment conditions, RADIUS updates the training set as discussed in Section \ref{sec:trainingSetUpdate}, dynamically generating new thresholds. For this, the relevant parameter is the update window size $l_{update}$. We first show that the detection performance is enhanced with this updating technique, and then we discuss the impact of $l_{update}$.

In Figure~\ref{fig:EVA-updateWindow}(a), we present an RSSI trace with a valley of RSSI values (between 300 and 500 seconds) indicating an abnormal situation that causes the monitored PDR to fall below the expected performance as shown in Figure~\ref{fig:EVA-updateWindow}(b). By adapting the training set and consequently the threshold (see Figure~\ref{fig:EVA-updateWindow}(c)), we can see from Figure~\ref{fig:EVA-updateWindow}(d) that in this experiment, the detection error can be reduced of 3\%-4\% with updated thresholds, down from 18\% to 14\% approximately.


Furthermore, we also observe from Figure~\ref{fig:EVA-updateWindow}(d) that the impact of $l_{update}$ is not significant on the detection error
(we set $l_{update}$ to be 50 for our example deployment). However, a larger $l_{update}$ may require a longer time to fill up the window making the threshold update less responsive in some cases. With such setting of $l_{update}$, we observe the detection error can be reduced of 3\% to 8\% in all experiments. Considering that the total error rate in most of our experiments is less than 20\%, such amount of reduction in the error rate is significant.

\subsection{Refinement of the A Priori Probability}\label{sec:EVA-paramRefine}

In addition to updating the training data set, one other situation that requires to generate a new threshold is when the detection accuracy degrades with an increasing number of false alarms, indicating the need of updating the \textit{a priori} probability. As described in Section \ref{sec:prioriRefinement}, we consider the maximum number of consecutive false alarms $N_{alarm}$ and the adjustment step $\delta$ of $P(H_g)$. We quantify the effects of these parameters in Figure \ref{fig:EVA-paramRefine}, where we compare the detection accuracy with and without the refinement of the \textit{a priori} probability. Specifically, we show the detection performance with varying $N_{alarm}$ and $\delta$. We use the suggested values in the above sections for the other parameters.


From Figure \ref{fig:EVA-paramRefine}, we can see that a smaller $N_{alarm}$ can reduce FPR but it may also cause a significant increase in FNR due to over-adjustment. On the other hand, a larger $N_{alarm}$ makes the system conservative on the \textit{a priori} probability refinement and hence the refinement less effective. The optimal choice of  $N_{alarm}$ falls at the location where the total error rate is lowest. In addition, the choice of the parameter $\delta$ needs to consider a tradeoff: larger $\delta$ indicates a more effective adjustment but a higher risk of over-adjustment. In our example, we choose $N_{alarm} = 5$ and $\delta = 0.003$. With such parameter settings, the analysis of all data traces shows that based on the accuracy improvement achieved by the training set updating technique, refining $P(H_g)$  can further reduce the error rate in a range from 2\% to 5\%. 

\section{Implementation and Evaluation} \label{sec:imp&eva}

In the previous sections, we presented the RADIUS system design and analyzed the impact of its parameters on the detection performance. Based on these, we implemented the DA component of RADIUS for TelosB sensor platforms and the VCC for standard PCs. In this section, we detail our implementation and discuss the system overhead. At last, we show the evaluation results on the detection performance of the overall implemented system in an indoor testbed. 

\subsection{System Implementation} 

In this section, we first describe the implementation details of the two major RADIUS components: the DA and the VCC. We introduce the programming interface of the DA to show that it is easy to use for higher-layer services and applications, followed by the implementation details of the VCC and of the RADIUS IoT extension. 

\begin{figure}[h]
	\vspace{-0.3cm}
	\begin{lstlisting}[frame=single]  % Start your code-block
	
	interface DetectionAgent {
	command void configureDA(Struct_Param parameters);
	command error_t start_Training();
	command error_t stop_Training();
	command error_t start_Detection();
	command error_t stop_Detection();
	command void update_RSSI(uint8_t rssi, uint8_t childId);
	}
	\end{lstlisting}
	\vspace{-0.55cm}
	\caption{\textbf{The programming interface for DA.}}
	\label{fig:TinyOSinterface}
	\vspace{-0.3cm}
\end{figure}

To ease the integration of RADIUS into higher-layer applications, we implemented the DA component as a module on TinyOS 2.1.2, which provides an interface \texttt{DetectionAgent} (see Figure \ref{fig:TinyOSinterface}). The \texttt{configureDA} command is used to configure DA with user-specified parameter settings (e.g., the initial $P(H_g)$) as provided in configuration messages sent by the VCC. The interface also provides control commands such as \texttt{start\_Training} or \texttt{start\_Detection} for executing the different phases in RADIUS. The command \texttt{update\_RSSI} is used to update the RSSI distribution of each link during both the training and the detection phases. With this programming interface, an application only needs to react to VCC's control messages and call the different commands accordingly. Note that RADIUS is not restricted to TelosB or TinyOS and can be easily adapted to other embedded devices with low-power radios.

\begin{figure}[t]
	\centering
	\includegraphics[width=1\linewidth, height = 5.5cm]{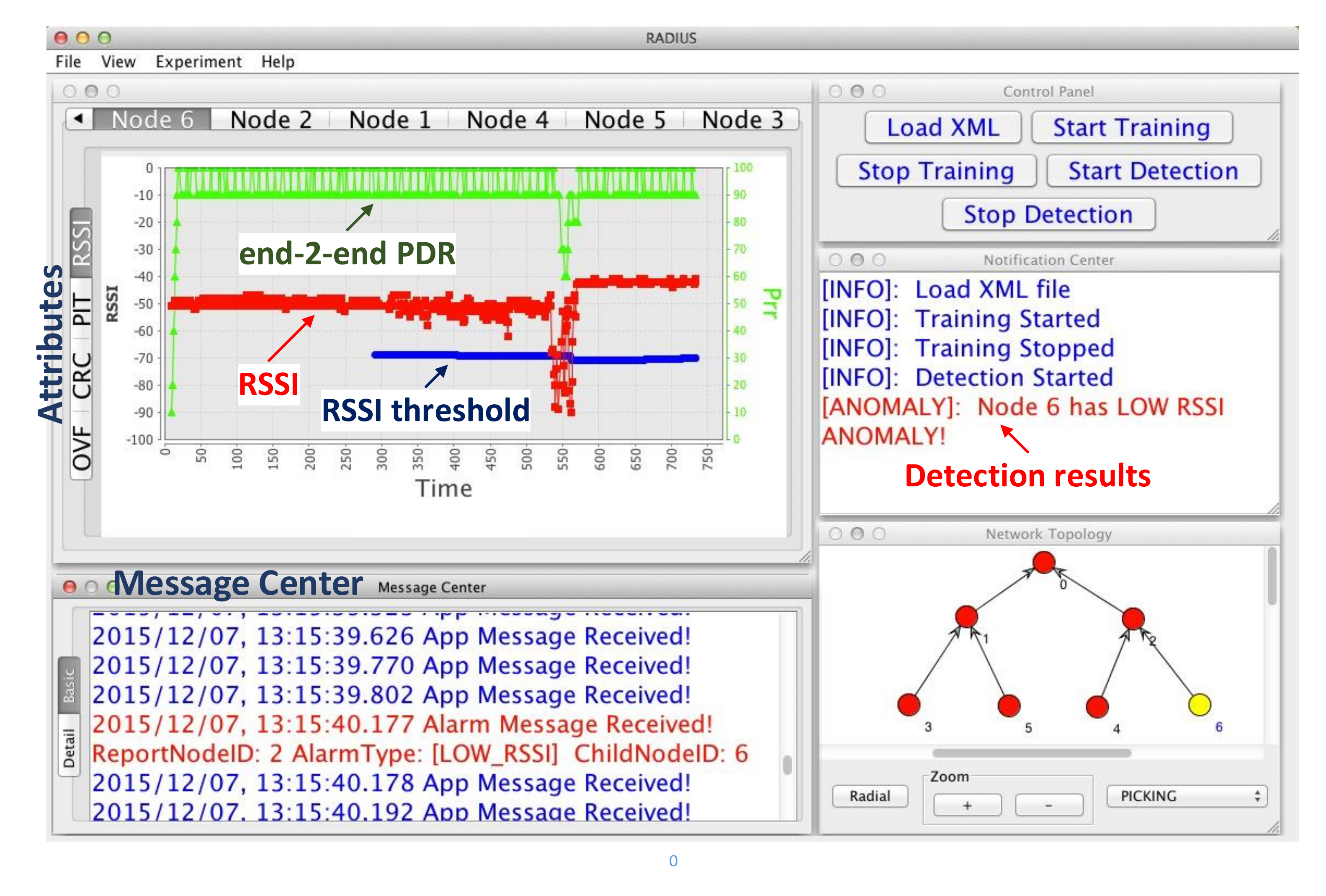}
	\vspace{-0.9cm}
	\caption{\textbf{The Monitoring User Interface of VCC.}}
	\label{fig:GUI}
	\vspace{-0.7cm}
\end{figure}

\begin{figure}[t]
	\centering
	\includegraphics[width=1\linewidth, height = 5.5cm]{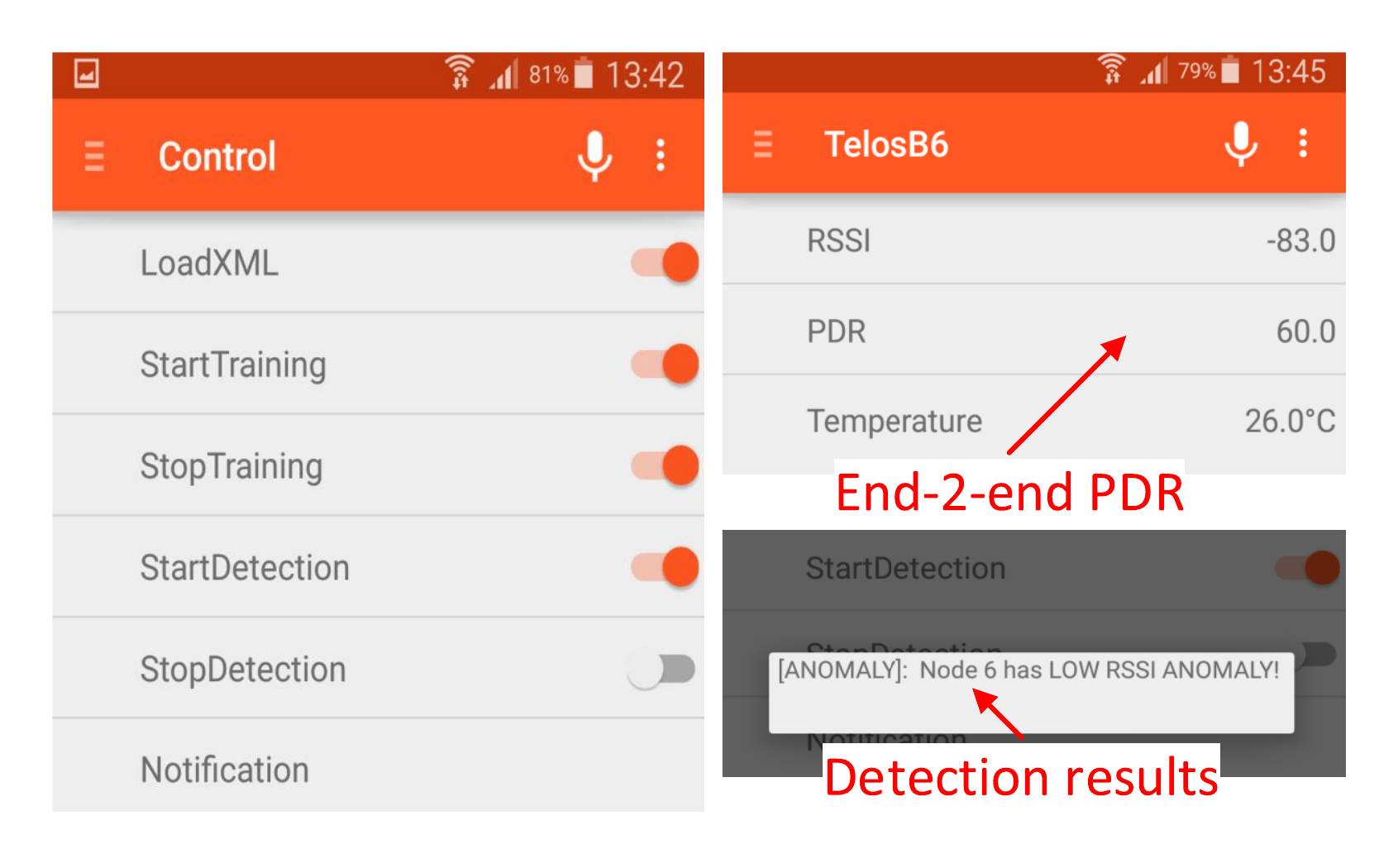}
	\vspace{-0.9cm}
	\caption{\textbf{The Android UI for RADIUS.}}
	\label{fig:IOT}
	\vspace{-0.7cm}
\end{figure}

The component VCC, running on a standard PC, is implemented in Java. It processes the alarms received from the network and produces diagnosis results reporting the relevant anomalies and their locations for the nodes that are experiencing high packet losses, which assists the system operator in identifying possible remedy actions. The VCC also includes a \textit{ Monitoring User Interface} (see Figure \ref{fig:GUI}), which provides a visualization of the packet delivery performance, detection status and the diagnosis results. Via this interface, the operator can monitor and control the RADIUS system.

We have also extended the VCC with an Internet-of-Things interface, which allows the VCC to connect with openHAB \cite{openHAB}, an open source smart home automation software. MQTT \cite{MQTT}, a lightweight messaging transport protocol, is used for the communication between openHAB and the VCC. With this extension, the user can remotely control the RADIUS system and monitor the detection results with an Android smart phone, as depicted in Figure \ref{fig:IOT}.

The current implementation of RADIUS is able to detect anomalies in link quality. Nevertheless, the programming interface (Figure \ref{fig:TinyOSinterface}) and the DA module can be easily extended to detect anomalies of other attributes, e.g., CRC error rate or packet overflow rate.
In addition, RADIUS currently works for static tree-based data collection applications but it can be also applied to other routing schemes. 

\subsection{System Overheads} \label{sec:overhead}
In this section, we analyze the memory, communication and computation overheads of our RADIUS implementation.


\paragraph{Memory overhead.} Detection Agents incur memory overhead on RAM (data) and ROM (program) of sensor nodes. As presented in Section \ref{sec:trainingSetUpdate}, we keep updating the training set to adapt the mean $\mu$ and standard deviation $\sigma$ of the density distribution of RSSI. To avoid increasing RAM usage during the update, we implemented this in a memory-friendly way, i.e., to compute $\mu$ and $\sigma$ with a single pass without storing the previous measurements of RSSI. To do so, we reformulate $\mu$  and $\sigma$ in the following way:
\setlength{\belowdisplayskip}{2pt} \setlength{\belowdisplayshortskip}{2pt}
\setlength{\abovedisplayskip}{2pt} \setlength{\abovedisplayshortskip}{2pt}
\begin{equation} \label{equ:stdComputation}
\mu = \frac{s}{n},   \quad  \sigma = \sqrt{\frac{1}{n-1}\big(q - \frac{s^2}{n}\big)}
\end{equation}
where $s$ and $q$ are defined as follows:
\begin{align}
s= \displaystyle \sum_i^n{x_i},\quad  q = \displaystyle \sum_i^n{x_i^2}
\end{align}
in which $x_i$ is the i-th RSSI reading. Instead of storing the entire training set, the DA then stores only 2 counters ($s$ and $q$) for each link to compute and update $\mu$ and $\sigma$. By doing so, the RAM consumption of RADIUS has a complexity of $O(mn)$, where $m$ is the number of links from direct child nodes and $n$ is the number of monitored attributes (e.g., RSSI, CRC error rate), remaining independent from the sample number.

To evaluate the RAM and ROM overhead, we compare the memory usage of a tree-based data collection application with and without the DA module. In the application, each node has two one-hop child nodes and therefore stores information about two links. The application alone consumes 3060 bytes in RAM and 25082 bytes in ROM while the application including the DA module consumes 3176 bytes in RAM and 31170 bytes in ROM. This indicates that the DA module consumes 116 bytes RAM (in comparison to 10 KB RAM in a TelosB device), and approximately 6 KB ROM (in comparison to 48 KB ROM in a TelosB device). 

\paragraph{Communication overhead.} Due to its distributed architecture, the anomaly detection process alone incurs no communication overhead in RADIUS. It requires additional communication only if the DAs send alarms corresponding to detected anomalies, or when the VCC sends control messages. To reduce such overhead, the alarms with minimum information (2 bytes) about the detected anomaly are piggybacked on the application packets. On the other hand, the number of control messages delivered from the VCC to the DAs, based on our indoor testbed evaluation results, is negligible compared to the amount of received application packets.

\paragraph{Computation overhead.} The main computation overhead comes from the processing of the Bayesian thresholding. In RADIUS, the complexity of Bayesian thresholding involves the calculation of the mean, the standard deviation and the Bayes threshold according to Equation \ref{equ:rssiTHD}. Testing results show that the processing of a Bayes threshold takes about 10 ms, which is small compared to the normal packet inter-arrival time in typical data collection applications.

\begin{figure}[t]
	\centering
	\includegraphics[width=1\linewidth, height=4cm]{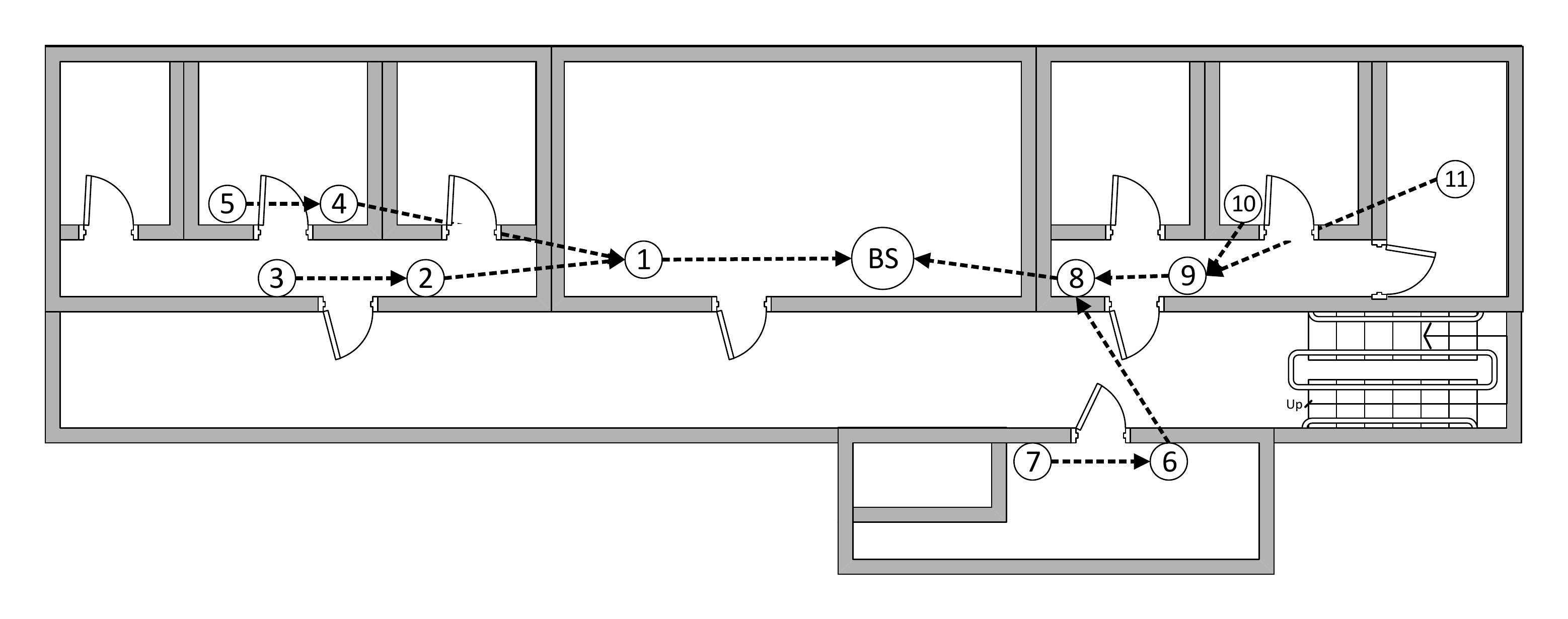}
	\vspace{-0.85cm}
	\caption{\textbf{The indoor testbed.}}
	\label{fig:floor-plan}
	\vspace{-0.35cm}
\end{figure}

\subsection{Experimental Evaluation} \label{sec:evaluation}

We have evaluated the detection performance of our RADIUS implementation in an indoor testbed (Figure \ref{fig:floor-plan}) consisting of 12 TelosB motes, deployed in several offices of a university building. Each sensor node runs an application that collects environmental data and sends it to the sink every 2 seconds following a tree-based routing topology. We instrumented a Detection Agent on each sensor node and ran the VCC on a PC connected to the sink node (Base Station).

\begin{figure}[t]
	\centering
	\includegraphics[width=1.0\linewidth, height = 7.5cm]{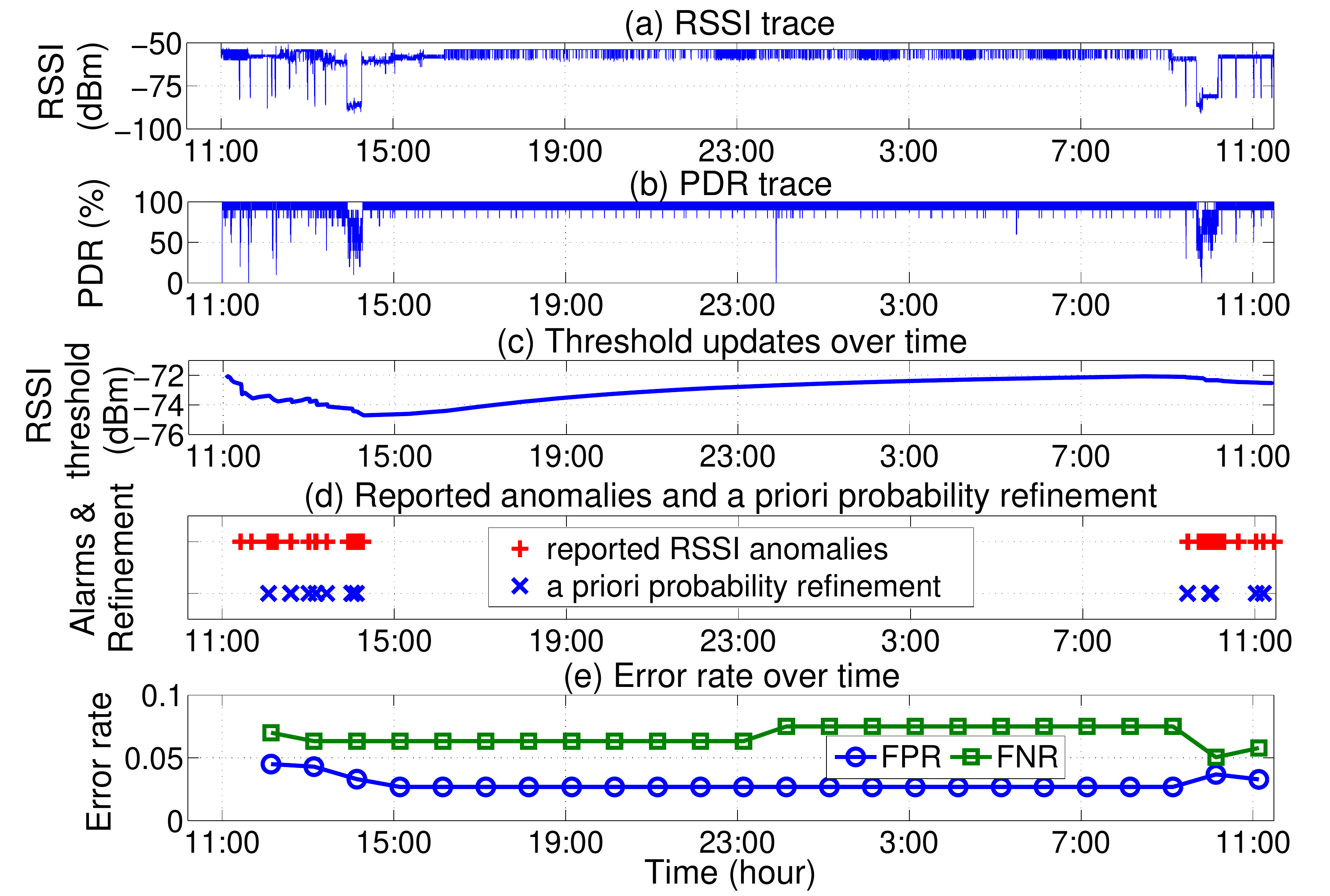}
	\vspace{-0.7cm}
	\caption{\textbf{Experimental results for link quality anomaly detection of the link from node 11 to node 9. The  experiment was run for 24 hours, from 11:00 am till 11:00 am on the next day.}}
	\label{fig:EVA-1node}
	\vspace{-0.2cm}
\end{figure}

To configure the RADIUS system, we adopt the system parameter settings suggested in the previous analysis (Section \ref{sec:parameterChoice}). Table \ref{tab:parameter} summarizes the suggested parameter settings. After a training period of about 5 minutes, RADIUS starts the detection phase for a period of about 24 hours. During the experiment, we logged the received alarms, the RSSI traces and the PDR traces. Figure \ref{fig:EVA-1node} demonstrates our results of detecting link quality anomalies on one of the links.

\begin{table}[t]
	\centering
	\caption{\textbf{The system parameter settings used in the evaluation.}}
	\vspace{-0.3cm}
        \footnotesize
	\begin{tabulary}{0.95\textwidth}{|l|l|l|}
		\hline
		\textbf{Techniques} & \textbf{Parameters} & \textbf{Settings} \\ 
		\hline
		\textbf{Bayesian} & initial $P(H_g)$ & 0.8\\
		\textbf{Thresholding} & maximum $P(H_g)$ & 0.99\\
		\hline
		\textbf{Training Set Size} & sample number $N_s$ to compute $\sigma_s$ & 250 \\
		\textbf{Estimation} & max. estimated error of mean $E_{\mu} $ & 1 dBm\\
		\hline
		\textbf{Data Smoothing}	& sliding window size $l$ & 3 \\
		\hline
		\textbf{Training Set Update}	& update window size $l_{update}$ & 50 \\ 
		\hline
		\textbf{\textit{A Priori} Probability} & max. alarm number $N_{alarm}$ & 5 \\ 
		\textbf{Refinement} & adjustment step $\delta$ & 0.003 \\ 
		\hline
	\end{tabulary}
	\vspace{-0.6cm}
	\label{tab:parameter} 
\end{table}

From the PDR trace depicted in Figure~\ref{fig:EVA-1node}(b), we can observe that the link frequently experienced high packet losses during the first 4 hours and the last 3 hours due to the bad channel quality caused by the students' movements crossing the communication link (see Figure~\ref{fig:EVA-1node}(a)). The received alarms that reported such anomalous link quality degradation are marked in red in Figure~\ref{fig:EVA-1node}(d). The results show a good detection accuracy. The overall FNR and FPR of detecting the anomalous RSSI degradation for this link over 24 hours are 5.1\% and 4\%, respectively. 

In addition, we can see from Figure~\ref{fig:EVA-1node}(e) that RADIUS can keep the error rate stable over the detection period. Figure~\ref{fig:EVA-1node}(c) clearly shows that the threshold is adaptive to RSSI variations due to the environment changes. From our analysis based on the logged refinement points (marked in Figure~\ref{fig:EVA-1node}(d)), RADIUS refines the \textit{a priori} probability at around 13:00 on the first day and 10:00 on the next day to reduce FPR and thus maintain the detection accuracy. 

\begin{figure}[t]
	\centering
	\includegraphics[width=1.0\linewidth]{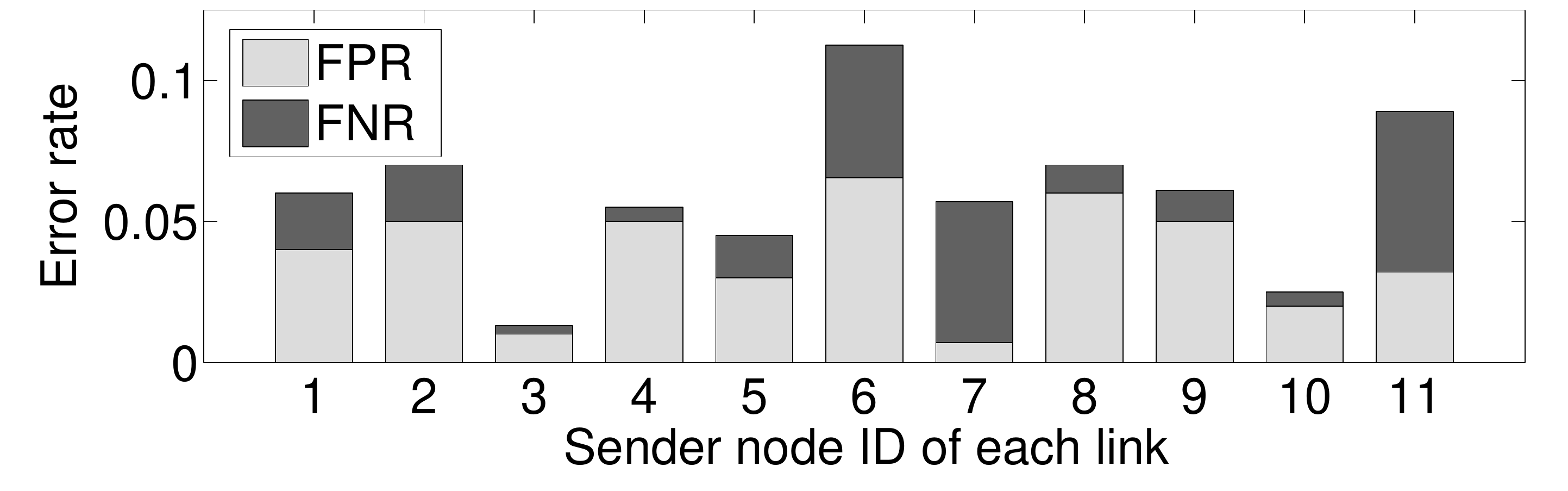}
	\vspace{-0.7cm}
	\caption{\textbf{The error rates for every link in the network (Figure \ref{fig:floor-plan}).}}
	\label{fig:EVA-11nodes}
	\vspace{-0.55cm}
\end{figure}

To demonstrate that RADIUS can robustly achieve high detection accuracy for all the links across the entire deployed network, we plot in Figure \ref{fig:EVA-11nodes} the error rate for every link in the testbed. The figure shows that with a set of global parameter settings (listed in Table \ref{tab:parameter}), RADIUS achieves a low error rate for every link in the network (6.13\% on average). 


\section{Conclusion} \label{sec:conclusion}

This paper presents RADIUS, a system for detecting anomalous link quality degradations in low-power radio links. The RADIUS system is light-weight, accurate and robust to a diversity of link conditions and dynamic environment changes. To achieve this, RADIUS (1) lays its foundation on a Bayesian thresholding scheme, integrated with dedicated techniques for (2) minimal training set size estimation, (3) sliding-window data smoothing, (4) distribution self-adaptation, and (5) feedback-based threshold parameter adaptation. The comparison with two popular statistical approaches shows that RADIUS does not need fine-tuning of its threshold parameter to achieve near-optimal accuracy across the network. The impact of the system parameters is also investigated in detail, identifying the best configuration for an indoor environment. Moreover, we have implemented the RADIUS system and evaluated its performance on an indoor WSN testbed, showing that it can adapt to dynamic environment changes and achieve accurate detection over the entire network with an average error rate of 6.13\%.

\balance
\bibliographystyle{abbrv}
\bibliography{IEEEabrv,AnomalyDetectionPaper}

\end{document}